\documentclass[a4paper,11pt,headings=big,DIV=12]{scrartcl}
\pdfoutput=1
\usepackage{amsmath,amssymb,mathtools}
\usepackage[table]{xcolor}
\usepackage{rotating}
\usepackage{tcolorbox}
\definecolor{mygray}{gray}{0.2}
 \addtokomafont{disposition}{\rmfamily\boldmath}
  \usepackage{bbold}
   \usepackage{environ}
 \renewcommand*{\arraystretch}{2.0}

\NewEnviron{thincases}{\scalebox{0.5}[1]{$\displaystyle
\left\{\scalebox{2}[1]{\setlength{\arraycolsep}{6pt}% <- I did not look up the "correct" value
$\displaystyle\begin{array}{ll}
\BODY
\end{array}$}\right.$}}%}

\definecolor{mypink1}{rgb}{0.9, 0.2, 0.6}
\usepackage{graphicx}
\addtokomafont{disposition}{\rmfamily\boldmath}
\usepackage[utf8]{inputenc}
\usepackage{enumerate}
\usepackage{slashed}
\usepackage{cite}
\usepackage{comment}
\usepackage{multirow}
\usepackage{acronym}
 \usepackage{bbold}
 \usepackage{tikz-feynman}
 \usepackage{jheppub}
\allowdisplaybreaks
\newcommand{\rr}{\rho_{\sig\pi}}
\newcommand{\rsES}{r^2_{E_0}}
\newcommand{\rsMS}{r^2_{M_0}}
\newcommand{\rsET}{r^2_{E_1}}
\newcommand{\rsMT}{r^2_{M_1}}
\newcommand{\rsA}{r^2_{A}}
\newcommand{\Isky}{I_{sky}}
\newcommand{\Lagsky}{\Lag_{\text{SD}}}
\newcommand{\MM}{M_{\text{sky}}}
\newcommand{\hXX}{{\hat{\chi}}}

\definecolor{mypink1}{rgb}{0.9, 0.2, 0.6}
\usepackage{graphicx}
\addtokomafont{disposition}{\rmfamily\boldmath}
\usepackage[utf8]{inputenc}
\usepackage{enumerate}
\usepackage{slashed}
\usepackage{cite}
\usepackage{comment}
\usepackage{multirow}
\usepackage{acronym}
 \usepackage{bbold}
\allowdisplaybreaks
\numberwithin{equation}{section}

\newcommand{\vev}[1]{\langle #1 \rangle}

\newcommand{\matel}[3]{\langle #1|#2|#3\rangle}

\newcommand{\al}{\alpha}
\newcommand{\be}{\beta}
\newcommand{\ga}{\gamma}
\newcommand{\Ga}{\Gamma}
\newcommand{\de}{\delta}
\newcommand{\De}{\Delta}
\newcommand{\la}{\lambda}

\newcommand{\eps}{\epsilon}

\newcommand{\sig}{ \sigma}

\newcommand{\om}{\omega}
\newcommand{\Om}{\Omega}

\newcommand{\GeV}{\,\mbox{GeV}}
\newcommand{\MeV}{\,\mbox{MeV}}

\newcommand{\EQ}{Eq.~}
\newcommand{\EQs}{Eqs.~}
\newcommand{\TAB}{Tab.~}

\newcommand{\FIG}{Fig~}

\newcommand{\SEC}{section~}
\newcommand{\SECs}{sections~}
\newcommand{\APP}{appendix~}

\newcommand{\REF}{Ref.~}
\newcommand{\REFs}{Refs.~}

\newcommand{\Lag}{{\cal L}}

\newcommand{\dphi}{d_\varphi}

\newcommand{\gone}{A}
\newcommand{\gtwo}{ D}

\newcommand{\gJ}{J}
\newcommand{\ferm}{N}

\newcommand{\Lageff}{\Lag_{\text{eff}} }

\newcommand{\XX}{\chi}

\newcommand{\gast}{\ga_*}

\newcommand{\bestp}{\be'_*}

\newcommand{\Tr}{\mathrm{Tr}}

\newcommand{\dof}{\text{dof}}

\newcommand{\fm}{\,\mbox{fm}}
\newcommand{\ORD}{{\cal O}}

\newcommand{\dbar}{{d-1}}
\newcommand{\Tud}[2]{T^{#1}_{\;\; #2}}
\newcommand{\TEMT}{ \Tud{\rho}{\rho} }
\newcommand{\veci}[1]{\underline{#1}}  %{{\mathbf{#1}}}

\newcommand{\mink}{\eta}

\definecolor{violet}{rgb}{0.94, 0.2, 0.8}%lightcoral
\definecolor{lightblue}{rgb}{0.39, 0.58, 0.93} %cornflowerblue
\definecolor{lightgreen}{rgb}{0.1, 0.73, 0.33}

\DeclareOldFontCommand{\tt}{\normalfont\ttfamily}{\mathtt}
\DeclareOldFontCommand{\bf}{\normalfont\bfseries}{\mathbf}
\DeclareOldFontCommand{\rm}{\normalfont\rmfamily}{\mathrm}

\begin{document}

\title{\boldmath The Skyrme-Dilaton and   Nucleon Observables}

\author[1]{Thomas Fischbacher,}
\author[2]{Roman Zwicky,}
\affiliation[1]{Google Research
Brandschenkestrasse 110, 8002 Z\"urich, Switzerland}
\affiliation[2]{Higgs Centre for Theoretical Physics, School of Physics and
Astronomy, \\
The University of Edinburgh,  Peter Guthrie Tait Road, Edinburgh EH9 3FD, Scotland, UK}

\emailAdd{tfish@google.com}
\emailAdd{roman.zwicky@ed.ac.uk}

\abstract{
We extend the Skyrme model by incorporating the $\sigma/f_0(500)$-meson as a dilaton. The motivation is twofold: the $\sigma$-meson is the lightest scalar resonance in QCD and is therefore the natural extensions to the pions of the classical Skyrme model, while its interpretation as a pseudo-Goldstone boson of spontaneously broken scale symmetry is supported by recent studies of gravitational form factors and related developments. Although the $\sigma$-meson is a broad resonance, for static observables it can be treated as real particle since its line-shape  is not expected to be essential. A combined fit to static nucleon observables yields $\chi^2/\mathrm{dof} $ close to unity, with substantial improvement over the original Skyrme model, while simultaneously reproducing the physical value of $F_\pi$, the overall scale of the model. }

\maketitle

\setcounter{page}{1}

\mdseries

\section{Introduction}

Since its proposal \cite{Skyrme:1961vq,Skyrme:1962vh}, its formulation in QCD \cite{Witten:1979kh,Balachandran:1982dw,Adkins:1983ya} (reviewed in \cite{Donoghue:1992dd,Ma:2016npf}), and its connection to fermion spin \cite{Finkelstein:1968hy,Witten:1979kh}, the Skyrme model has found wide-ranging applications \cite{Brown:2010api} to nucleon and nuclear properties \cite{Manton:2004tk,Manton:2022fcb}. Its relation to the large-$N_c$ expansion provides a qualitative justification for treating the baryon as a classical soliton \cite{Witten:1979kh}. Nevertheless, the minimal Skyrme model has well-known phenomenological limitations: when confronted with nucleon observables, its natural scale, set by the pion decay constant, falls short by about thirty percent, and many observables deviate by a similar amount. It has been suggested that the large-$N_c$ structure alone cannot account for these deviations, and that improvement should come from adding further mesons \cite{Witten:1984cg}, in line with the idea that the baryonic sector emerges from the mesonic one.

In this spirit, many authors have added the lowest-lying vector mesons, the $\omega$ and/or the $\rho$, motivated by their phenomenological relevance in the vector-meson-dominance approach \cite{Adkins:1983nw,Meissner:1986vu,Meissner:1986ka,Meissner:1986js,Sutcliffe:2008sk,Ma:2012kb,Naya:2018mpt}. The lowest-lying resonance in QCD, however, is the $f_0(500)/\sigma$-meson. It is qualitatively different from the pions 
and the vector mesons: it is very broad $\sqrt{s_\sig} = m_\sig - \frac{i}{2} \Ga_\sig \approx (440 - \frac{i}{2} 570) \MeV$  \cite{ParticleDataGroup:2026mpi} and has been in and out of the 
Particle Data Group (PDG)  listings over the years  \cite{Pelaez:2015qba}.\footnote{The $\sigma$ is sometimes viewed as a dynamically generated resonance that should not appear in an effective Lagrangian (see, the discussion in the review  \cite{Pelaez:2015qba}). Here we pursue the opposite viewpoint.} For static (Euclidean) observables considered here, its effects can nevertheless be represented by an effective scalar degree of freedom with a characteristic mass $m_\sigma \approx 550(50)\,\MeV$, as used in fits to gravitational form factors \cite{Stegeman:2025sca,Stegeman:2025tdl} (and references
therein).

Independent motivation for treating the $\sigma$ as a genuine dynamical state, rather than a mere unitarisation object, has accumulated recently from two directions. On the lattice, studies of gauge theories near the edge of the conformal window consistently reveal a light flavour-singlet scalar (see \cite{Fodor:2017gtj,LatticeStrongDynamics:2023bqp,LatKMI:2025kti} for recent works). In parallel, analytic work built on the hypothesis that QCD itself lies close to an infrared fixed point has identified the $\sig$ with the dilaton, the (pseudo-)Goldstone boson of spontaneously broken scale symmetry \cite{Crewther:2013vea,Crewther:2015dpa,Zwicky:2023bzk,Zwicky:2023krx}. This identification has now been confronted directly with lattice data on gravitational form factors \cite{Stegeman:2025sca,Stegeman:2025tdl}: the residues extracted from a $\sigma$-pole fit agree with the predictions of dilaton effective theory, lending concrete, quantitative support to the picture that the $\sigma$ couples like a dilaton rather than like an ordinary phenomenological meson. This convergence of lattice and analytic evidence provides a symmetry-based motivation for introducing the $\sigma$ into the Skyrme model on this footing.\footnote{The dilaton has appeared elsewhere such as describing  approximate scale symmetry in strongly interacting matter \cite{Brown:1991kk,Rho:2021zwm,Rho:2024jdo}, and in holographic models, where light dilaton-like states have been associated with either large condensates \cite{Elander:2017cle,Elander:2017hyr} or proximity to weak, second-order phase transitions \cite{Elander:2020ial,Faedo:2024zib,Elander:2025fpk}. A further line of investigation concerns the possibility that the Higgs boson itself is a dilaton \cite{Matsuzaki:2012xx,Dietrich:2005jn,Cata:2018wzl,Zwicky:2023krx}.}

In this work we therefore extend the Skyrme model by incorporating the $\sigma$-meson as a dilaton.\footnote{For a previous study using the so-called  log potential  and focusing on gravitational form factors see  \cite{Tanaka:2025pny}. 
The paper does not contain sufficient overlap for comparison.}  
The resulting 
Skyrme-dilaton theory combines the nonlinear realisation of chiral symmetry with a compensator field for Weyl transformations. The construction is based on the leading-order dilaton chiral effective Lagrangian, supplemented by the four-derivative Skyrme term. The improvement term required by local Weyl invariance is of particular importance: it fixes the coupling of the dilaton to the energy-momentum tensor and, consequently, the long-distance contribution of the scalar field to the gravitational form factor. In this way the soliton solution provides a concrete realisation of the connection between a dilaton and the nucleon's gravitational $D$-form factor.

The paper is organised as follows. 
In \SEC\ref{sec:SD}, the model is presented and the asymptotics of the $\sig$-field are derived.  
The numerical solutions are confronted against static observables in 
 \SEC\ref{sec:pred}.  
The paper ends with  summary and conclusions in \SEC\ref{sec:conc}.
Additional plots and computational details are deferred to appendices.

\section{The Skyrme-dilaton model}
\label{sec:SD}

In our model, the $\pi$ and  the $\sig$ are the Goldstone bosons of spontaneously broken chiral and scale symmetry  
\begin{equation}
U = \exp( 2 i \, \pi^a T^a /F_\pi) \;, \qquad \hXX = \exp( - \sig/F_\sig) \;. 
\end{equation}
They transform non-linearly under  chiral  $(L,R) \in  SU(2)_L \otimes SU(2)_R$  and  (local) Weyl $g_{\mu\nu} \to e^{-2 \al(x)} g_{\mu\nu} $  transformations 
\begin{equation}
(  U \to  L U R^\dagger, \, \hat{\chi} \to \hat{\chi} )_{\text{chiral}}  \;,  \qquad (U \to U , \,  \hat{\chi} \to e^{\al} \hat{\chi})_{ \text{Weyl}}  \; .
\end{equation}
Intuitively, the $U$-field does not transform under Weyl transformation since it is an angular variable. 
Throughout,  $\hXX$ will sometimes be referred to as the dilaton field.

The relevant Skyrme-dilaton (SD) Lagrangian is the leading order dilaton Lagrangian
in chiral perturbation theory ($\chi$PT)
\cite{Zwicky:2023fay,Zwicky:2023krx} 
%($\chi \equiv F_\sig \hXX$)
\begin{alignat}{2}
\label{eq:Leff}
 & \Lagsky   &\; =  \;&   \frac{ F_\pi^2 }{4}  \hXX^{2}  \Tr [ \partial  U \cdot \partial  U^{\dagger}] +   
 \Lag_{m_\pi} 
    \nonumber   \\[0.1cm]
&     &\; +  \;&  \frac{F_\sig^2}{2}\left(  (\partial \hXX)^2  - \frac{R}{6} \,  \hXX^{2} \right)   + \frac{1}{32 e^2}  
\Tr ( [ R_\mu,  R_\nu  ])^2    \;,
\end{alignat}
augmented by the Skyrme-term which can be conveniently  written in terms of the (Lie algebra-valued) Maurer-Cartan 
form $R_\mu = ( \partial_\mu U)  U^\dagger$.  The pion mass term  for $N_f=2$ mass-degenerate quarks  is given by 
\begin{equation}
\label{eq:Lmpi}
\Lag_{m_\pi} =  \frac{m_\pi^2 F_\pi^2}{4}  ( \hXX^2  \Tr[ U^\dagger + U ]   - 2 \hXX^4 - 2) \;,
\end{equation}
and includes the $\sig$-potential.  Expanding to second order in $\sig$ uncovers  its mass 
\begin{equation}
\label{eq:msig}
m_\sig = \frac{2 m_\pi}{\rr} \;, \quad \rr \equiv  \frac{F_\sig}{F_\pi} \;,
\end{equation}
where $F_{\sig,\pi}$ will both be discussed shortly. 
The  term proportional to the identity  in \eqref{eq:Lmpi}  implements the  subtraction of  the disconnected part
(vacuum expectation value), thereby ensuring a  finite energy solution. 
In the limit $\hXX \to 1$, which  will be referred to as the \emph{Skyrme-limit},  one recovers the standard Skyrme Lagrangian \cite{Adkins:1983ya}.\footnote{\label{foot:Fpi} In this reference a different convention for the pion decay constant is used, $F_\pi^{\text{ANW}} = 2 F_\pi$, which affects formulae for observables.}   
The Wess-Zumino-Witten (WZW) term, being a 5-form, vanishes for $SU(2)$ as it is a $3$-manifold only, and is therefore absent.

The construction principle of the Lagrangian \eqref{eq:Leff} is invariance under the two symmetries discussed above. 
The $\chi$-field acts as a compensator to implement local Weyl invariance. Since the Skyrme term has four derivatives,
there is no $\chi$-field coupling to it.  
 The term proportional to the Ricci-scalar $R$ (improvement term) ensures local Weyl invariance together 
 with the dilaton kinetic term. It is crucial as it implements the Goldstone matrix element
 \begin{equation}
\label{eq:Goldstone}
\matel{\sig(q)}{T_{\mu\nu}}{0} =  
\frac{F_{\sig} }{3}  
(m_{\sig}^2 \mink_{\mu\nu} - q_\mu q_\nu) \;,
\end{equation} 
which defines the dilaton decay constant, the order parameter of spontaneous scale symmetry breaking 
analogous to $F_\pi \approx 93 \MeV$
\begin{equation}
\label{eq:fdecay}
  \matel{\pi^b(q)}{J_{5\mu}^a(x)}{0} = 
i F_\pi  q_\mu  \de^{ab}   \;,
\end{equation}
being the  order parameter of spontaneous chiral symmetry breaking. 

The effective Lagrangian \eqref{eq:Leff} relies on two results from matching to QCD 
in  a fixed-point interpretation: the quark-mass anomalous 
dimension $\gast = \ga_{m_q}|_{\mu=0} = 1$ \cite{Zwicky:2023bzk,Zwicky:2023krx}  at the fixed point and 
the vanishing of the $\be$-function slope  $\bestp = \be'|_{\mu=0}$  
\cite{Zwicky:2023krx} (see also \cite{Shifman:2023jqn}).  The former ensures that 
$ \hXX^{3- \gast}  \Tr[ U^\dagger + U ]$ assumes   $ \hXX^{2}  \Tr[ U^\dagger + U ]$ and the latter avoids 
further trace anomaly matching, as parametrised in \cite{Cata:2018wzl} for example. 

The Lagrangian \eqref{eq:Leff} introduces the  four parameters 
$( F_\pi  \;, e \;,  \hat m_\pi \;, \rr  ) $ where 
\begin{equation}
\label{eq:four}
  \hat{m}_\pi = \frac{m_\pi}{e F_\pi}  \;,  \qquad   \rr  \equiv \frac{F_\sig}{F_\pi} \;,
\end{equation}
are dimensionless variables with $\rr$ already defined in \eqref{eq:msig}.  The quantities $F_\pi$ and $m_\pi$ are experimentally  well-defined 
whereas $e$ is a low energy constant to be determined indirectly. 
To discuss $F_\sig$, we first need to clarify a few points. In the real world, the $\sig$-meson is unstable and has a sizeable width, which may raise the question of whether its inclusion as an explicit degree of freedom is justified. The key point is that we consider only static (Euclidean) quantities. For such observables, the $\sig$ can be effectively mimicked by a stable particle with a mass of $m_\sig \approx 550(50)\,\MeV$ (see \cite{Stegeman:2025sca} and references therein). From \eqref{eq:msig}, we see that this corresponds to $\rr \approx 0.5(1)$. Although $\rr$ is not known a priori, there are indications that it is not too far from unity. 
For a more comprehensive assessment of the $\sig$ as a dilaton across more channels, the inclusion of the strange-quark mass is presumably essential \cite{Crewther:2013vea,Zwicky:2023krx}. However, for the observables considered here, which do not involve isolated $0^{++}$-channels, we expect this effect to be less important. Our results support this expectation, with $\rr$ treated as a fit parameter.

We briefly digress to discuss Derrick's theorem \cite{Derrick:1964ww} in the context of the Skyrme model 
since it motivates the addition of the  Skyrme-term and provides a powerful numerical crosscheck.  
Let us first neglect the pion mass and rescale units by $x \to x/\la$. We then have 
\begin{equation}
E_{\text{tot}}(\la) = \la E_2 + \frac{1}{\la} E_4 \;,
\end{equation}
where $E_{2,4}$ are the two- and four-derivative terms respectively.  The parameter $\la$ governs the size 
of the soliton and stabilises when it is extremal 
 $\partial_\la E_{\text{tot}}(\la) |_{\la=1}=0$, implying $E_2 = E_4$. So, clearly if there is no Skyrme-term $E_4=0$, 
 then $E_2=0$, and there is no finite energy solution.
Adding pion mass, which corresponds 
 to a zero-derivative term, modifies  the relation  to  
\begin{equation}
\label{eq:power}
E_2 = E_4 - 3 E_0  \;,
\end{equation}
which is a powerful sum rule satisfied by minimised solutions.  In the Skyrme context, this is sometimes 
referred to as (an instance of) the virial theorem. 

Next one needs to minimise the energy which can be obtained from 
the  energy-momentum tensor (EMT) by metric variation  
$T_{\mu \nu} =2  \frac{ \de  }{  \de{g^{\mu\nu}}}  \sqrt{-g} \Lagsky|_{g_{\mu\nu} \to \eta_{\mu\nu}}$ 
($g$ denotes the metric determinant and
$\eta_{\mu\nu}$ is the Minkowski metric with mostly-minus signature). One obtains 
%\footnote{This EMT matches the one in (they have a potential)   \cite{Tanaka:2025pny}  \EQ 54 
%(which is the traceless part $\bar{\Theta}_{\mu\nu}$). In the trace part $\hat{\Theta}_{\mu\nu}$ they seem to have forgotten all the kinetic terms.}
 \begin{equation}
  T_{\mu\nu} = 
     \frac{F_\pi^{2}}{2}  \hXX^{2}   \Tr [ \partial_\mu  U \partial_\nu  U^{\dagger}] +   
     F_\sig^2(  \partial_\mu \hXX  \partial_\nu \hXX      + 
     \frac{ 1}{6}  (\eta_{\mu\nu} \partial^2 -  \partial_{\mu} \partial_{\nu} )      \hXX^{2}  ) 
         +  \de T_{\mu \nu}^{\text{sky}}    - \eta_{\mu\nu} \Lag   \;,
   \end{equation}
   with $\de T_{\mu \nu}^{\text{sky}} =     \frac{1}{8 e^2}  \Tr (  [R_\mu, R_\rho   ] 
       [ R_\nu , R^\rho ]) $.  The energy-density part is given by
\begin{equation}
T_{00}^{\text{stat}} =   \frac{ F_\pi^2 }{4}  \hXX^{2}  \Tr [ \partial^i  U \partial_i  U^{\dagger}] +   \frac{F_\sig^2}{2}(  (\partial_i \hXX)^2 
   - \frac{1}{3} \partial_i^2 \hXX^2) + \frac{1}{16 e^2} \Tr [ \tilde{R} \cdot (\tilde R)^\dagger ]    - \Lag_{m_\pi}   \;,
\end{equation}
where $\tilde{R}_k =   \eps_{ijk} R_i R_j $. 
In anticipation of the static ansatz, the static limit was assumed such that  
$ \partial_i^2 \equiv \veci{\partial}^2$ denotes a  3D Euclidean Laplacian (i.e. summation over indices  $i$ assumes a Euclidean metric $\de_{ij}$). 

\subsection{Hedgehog ansatz and equation of motion}

The associated Hamiltonian 
is obtained from  $H = \int d^3 x T_{00}$ and to make further  progress we now commit to the famous 
hedgehog ansatz for the skyrmion  
\begin{equation}
\label{eq:U0}
{U_0}(x) = \cos ( f(r) ) + i( \veci{ \tau} \cdot \veci{\hat{x}}) \, {\sin (f(r))}\;.
\end{equation}
which intertwines space and isospin rotations in a non-trivial manner since
  $\veci{\hat{x}}$ is the unit $x$-vector and $\veci{\tau}$ are the (isospin) Pauli matrices.
The profile function $f$, sometimes referred to as the skyrmion, is solely dependent on the radius. 
 The total energy is obtained by minimising 
\begin{equation}
\label{eq:M0}
\MM = \int d^3 x\, T_{00}^{\text{stat}}(f,\hXX)|_{\text{min}} = M_0 \left[\frac{F_\pi}{e}  \right]   \;, \quad
 M_0 =  2\pi \int_0^\infty  dr\, r^2 \rho_{M_0}(r)
\end{equation}
with mass density
\begin{equation}
\rho_{M_0}(r)  =   \hXX^2 (   f'^2 + 2 \frac{s^2}{r^2}) +  \rr^2 ( (\hXX')^{2} - \frac{1}{3} \veci{\partial}^2 \hXX^2 )+ 
 \frac{s^2}{r^2}(\frac{s^2}{r^2} + 2f'^2) + \hat{m}_\pi^2  (1+ \hXX^4  - 2 \hXX^2 c   )    \;,
\end{equation}
where primes denote derivatives with respect to $r$ and hereafter  
\begin{equation}
\label{eq:short}
s = \sin f \;, \qquad  c = \cos f \;,  \qquad s_2 = \sin 2 f \;.
\end{equation}
Moreover, we have silently redefined\footnote{One can introduce the dimensionless coordinate $\tilde r = a/r$ and express all integrals in terms of $\tilde r$ and the corresponding profile \(\tilde f(\tilde r)=f(r)\). In the following, we suppress the tildes from the outset and work in dimensionless units. Factors that restore the appropriate dimensions, such as the accompanying powers of $\hat m_\pi$ in \eqref{eq:four}, will be kept explicit where required by the context.
}  
\begin{equation}
\label{eq:a}
r \to  r a \;,    \qquad a \equiv \frac{1}{e F_\pi} \;,
\end{equation}
in order to work with a dimensionless variable $r$. 
 The scale assumes about half a Fermi,  $a  \approx 2.5 \GeV^{-1} \approx 0.5 \fm$, 
for characteristic references values of $F_\pi = 92.1 \MeV$ and $e = 4.35$. It sets the overall scale similar to the lattice spacing 
in Monte Carlo simulations.

The equations of motion for $M_0$ follow from variation by $f$
\begin{equation} 
\label{eq:f}
{
( \hXX^2  r^2 + 2 s^2 ) f'' + (2 r \hXX^2+   2 r^2  \hXX \hXX' ) f' +  s_2 f'^2 - 
( \hXX^2 + \frac{s^2}{r^2} )   s_2  - { r^2  \hat{m}_\pi^2 \hXX^2  s  }  = 0 } \;,
\end{equation}
% MS gets the f'-term completely wrong ... sets \hXX =1 and misses terms 
and  $\hXX$ 
\begin{equation}
\label{eq:chi}
{\hXX'' + \frac{2}{r} \hXX' - K_\pi(r)  \hXX = 0 }\;, 
\end{equation}
where
\begin{equation}
K_\pi(r) =  \frac{1}{ \rr^2}  (f'^2 + 2 \frac{s^2 }{r^2}   +     2 \hat{m}_\pi^2 ( \hXX^2  -  c  )    ) \;.
\end{equation}
% MS gets this equation correct though 
The expressions are consistent with the equation of motion in \cite{Zwicky:2023fay}. 
Note that since we are in the static case, minimisation of the energy is  equivalent to solving the equation of motion. 

\subsection{Rigid rotator quantisation}

The solutions $f$ and $\hXX$ are  spherically symmetric by the nature of their ansatz, but the nucleon is not as it has spin. 
This can partly be captured by quantisation of the rotational degrees of freedom \cite{Adkins:1983ya}, also known as moduli or collective coordinates, which we will review here for completeness and for introducing the role of the dilaton. 
  The starting point is the observation that for any  $A \in SU(2)$, 
$U' = A U_0  A^\dagger$ forms a new solution, and if $A(t)$ is taken to be time-dependent then one obtains 
\begin{equation}
\Lageff  = \Lagsky  + \Isky  \, \Tr[ \partial_0 A^\dagger \, \partial_0 A ] \;,
\end{equation}
where 
\begin{equation}
\label{eq:ISkyrme}
\Isky = I_0  \left[ \frac{1}{F_\pi e^3} \right]  \;, \quad I_0 = \frac{8\pi}{3}   \int_0^\infty dr \, r^2 s^2 \left(\hXX^2  +  
 (f'^2 +  \frac{s^2}{r^2} ) \right)  \;.
\end{equation}
In the Skyrme-limit $\hXX\to 1$, this agrees with  \EQ 4.64  \cite{Ma:2016npf} and 
the expression in  \cite{Adkins:1983ya} when taking  conventions  into account. 
Time-dependence allows for canonical  quantisation and gives the following correction \cite{Adkins:1983ya}
\begin{equation}
\label{eq:Mtot}
M_{\text{tot}} = \MM + \frac{1}{2 \Isky } j(j+1) \;,
\end{equation} 
upon using that isospin and angular momentum are intertwined $I^2 = J^2$. 
The quantity $M_{\text{tot}}$ is the mass after inclusion of the ``rigid-rotator" quantum corrections. 
The expression in \eqref{eq:Mtot} makes it clear that
 $ j = \frac{1}{2} , \frac{3}{2} $ correspond to  the nucleon and the $\De$-baryon respectively:
\begin{equation}
\label{eq:mNDe}
m_N = M_0\left[\frac{F_\pi}{e}\right] + \frac{3}{2} \frac{1}{I_0} \left[e^3 F_\pi\right], \qquad
m_\Delta = M_0\left[\frac{F_\pi}{e}\right] +  \frac{15}{2} \frac{1}{I_0}  \left[e^3 F_\pi\right] \;.
\end{equation}
We see that the correction to the nucleon mass is $\frac{1}{4} (m_\De - m_N) \approx 73\MeV$ corresponding to a  
$8\%$-correction. The quantisation procedure affects many other observables,  as we will see in \SEC\ref{sec:pred}.

The overall scale of the corrections is still set by the pion decay constant, but they involve different powers of $e$. This affects their large-$N_c$ behaviour, which is worth a digression  (see Appendix \ref{app:largeNc} for an extended discussion). It is well established that $F_\pi = \ORD(N_c^{1/2})$ and   $m_N = \ORD(N_c)$, implying \footnote{Alternatively, this scaling can be deduced by expressing $1/e^2$ as a linear combination of low-energy constants of chiral perturbation theory; see, e.g., \cite{Donoghue:1992dd}.} $e = \ORD(N_c^{-1/2})$. Consequently, $m_\De-m_N = \ORD(1/N_c)$, which amounts to relative suppression of $1/N_c^2$  of the quantum-versus-classical contribution.

\subsection{Boundary conditions}

The ordinary coupled differential equations of second order \eqref{eq:f} and \eqref{eq:chi}, 
require  two boundary conditions for each field. For the Skyrme field $f$, these are related to the 
baryon current 
\begin{equation}
B^\mu = \frac{1}{24 \pi^2} \eps^{\mu \al \be \ga} \Tr[ R_\al R_\be R_\ga] \;,
\end{equation} 
which is a 3-form and can be obtained by gauging the WZW-term with left-right sources $l_\mu = r_\mu = \frac{1}{3} B_\mu^{\text{ex}}$ such that $\Lag^{eff} \supset B^\mu  B^{ex}_\mu$. Other terms do not contribute since the 
pion and dilaton fields are of baryon number zero.  The dilaton field $\hXX$ does not alter  
$B^\mu$ since it is  a conserved current of scaling dimension three and does not require powers of $\hXX$ as compensators. 
The number 
\begin{equation}
 N_B = \int d^3 x \, B_0 =  n  \quad   \Leftrightarrow  \quad f(0) - f(\infty)    = n \pi \;,
\end{equation}
is an integer characterising the homotopy $\pi_3(SU(N_f)) = \mathbb{Z}$ and 
corresponds to the baryon number. 
The nucleon has baryon number $B=1$, and the conventional choice is 
\begin{equation}
f(0) = \pi \;, \qquad f(\infty) = 0 \;,
\end{equation}
which settles the $f$-boundary conditions but one may go further and investigate the asymptotic behaviour.
Assuming that the dilaton field is constant at both boundaries (to be confirmed), the same 
analysis applies as in the Skyrme model:
 demanding finiteness towards the $r \to 0$ integration for $M_0$ \eqref{eq:M0} and by linearising for $r \to \infty$ it
 follows that
 \begin{equation}
\label{eq:asym}
f(r)|_{r \to 0} = \pi - \omega \, r \;, \qquad f(r)|_{r \to \infty}  =  B (1 +\hat{m}_\pi r)\frac{   e^{-\hat{m}_\pi r }}{r^2}  \;,
\end{equation}
where $\omega$ is unknown and 
$B$ is related to the Goldberger-Treiman relation  \cite{Adkins:1983ya}, see the discussion in \APP\ref{app:con},  as we will do in the dilaton case in the next section.

  We may then turn to the dilaton field.  At large distances, \eqref{eq:asym} 
implies  that we may neglect  $K_\pi = \ORD(r^{-4})$ and   from \eqref{eq:chi} one finds the solution 
\begin{equation}
\label{eq:asymlarge}
\hXX(r)|_{r \to \infty} = A'  -  A \frac{ e^{-\hat{m}_\sig r }}{r}  \;,
\end{equation}
with a priori unknown constants $A$ and $A'$, and the minus sign chosen in anticipation of the result. 
At short distances  $f = \pi - \omega r$ and thus 
$K_\pi =  3    \omega^2 /\rr^2 = \Om^2 > 0$ (and let us use $\Om >0$ as well) for which the equation solves to 
\begin{equation}
\label{eq:asymshort0}
\hXX =  \frac{1}{r} (c  \sinh (\Om r)   + c' \cosh (\Om r) )  \;.
\end{equation}
Now, since we have a $dr \, r^2 (\hXX')^2$-term in the $M_0$-integral, finiteness demands $c'=0$
and thus
\begin{equation}
\label{eq:asymshort}
\hXX(r)|_{r \to 0}  =   c ( \Om    + \frac{1}{3!} \Om^3 r^2 +  \ORD(r^4))    \;, 
\end{equation}
where we recall that $\Om = \sqrt{3} \omega/\rr $ and  $c$  an unknown  constant.

Let us return to the question of whether the $f$-  \eqref{eq:asym} and  $\hXX$-asymptotics    are self-consistent. 
It seems so.  At short distances  the $f''$-term drops 
and we note that for $ f= \pi - \omega r $, out of the remaining four terms
the $\chi^2$-terms cancel separately in \eqref{eq:f}. At large distances, constant $\chi^2$-terms are leading as per \eqref{eq:asymlarge}  and therefore equivalent to the standard Skyrme analysis. 

%\paragraph{In the computer simulation}

For the actual numerical calculation, concrete boundary conditions are needed.  
In order to obtain a non-trivial solution we must impose a finite value at infinity. We may choose 
$\hXX(\infty) = 1$ which is natural if we assume that the $\sig$-field vanishes at infinity. 
This is not arbitrary, since
 formally one can always rescale (cf. \eqref{eq:Leff}) by 
 \begin{equation}
 \label{eq:rescale}
 ( \hXX ,  e^{-1} ,  F_\sig )  \to   e^{\al} ( \hXX ,  e^{-1} ,  F_\sig ) \;,
 \end{equation}
and this leads to an overall factor of $e^{2 \al}$ in front of the action.  Thus one may always absorb 
the value at infinity by other parameters. 
%Equivalently, and possible more transparent, one
 %may absorb it into a dilation transformation. 
Moreover, finiteness of the energy automatically enforces $\hXX = \text{const}$, consistent with our short 
distance analysis \eqref{eq:asymshort}.  So, in summary we have 
\begin{equation}
\label{eq:B}
\hXX(0) = \text{const}  \;,  \qquad  \hXX(\infty) = 1 - A \frac{ e^{- \hat{m}_\sig r}}{r} \;,
\end{equation}
where we will see in the next section that $A$ is determined by the dilaton Goldberger-Treiman relation 
ensuring the conformal Ward identity.

\subsection{The $\sig$-asymptotics and the conformal Ward identity } 
\label{sec:GT}

The key to satisfying the conformal Ward identity $\TEMT =0$ and retaining a finite nucleon mass is that the 
gravitational $D$-form factor is to assume the form (see \APP\ref{app:grav})
\begin{equation}
\label{eq:rN}
D(q^2) = \frac{r^N}{q^2}  + \ORD(1) \;,\qquad r^N  = \frac{4}{d-1} m_N^2 \;,
\end{equation}
which is realised by the dilaton Goldberger-Treiman relation \cite{Gell-Mann:1969rik,Carruthers:1971vz,DelDebbio:2021xwu,Zwicky:2023fay}
\begin{equation}
\label{eq:dGT}
g_{\sig NN} = \frac{m_N}{F_\sig} \;. 
\end{equation}
The central point is that in the spontaneously broken phase, the Ward identities are still intact, as is the 
case in the chirally broken phase where   the pion  Goldberger-Treiman relation \cite{Goldberger:1958tr}  ensures the chiral Ward identity. In the Skyrme model, $g_{\pi NN}$ governs  the pion far-field \cite{Adkins:1983ya} and
we aim to do  the same for the $\sig$ far-field in relation to $g_{\sig NN}$.  
Our presentation is logically different but equivalent  to  \cite{Adkins:1983ya}, which is sketched in \APP\ref{app:con}.

In order to carefully track factors, we consider it worthwhile to work in $d$ dimensions in this section. 
Now,  the $\sig$-cloud of the nucleon is determined by the interaction 
   $\Lageff = g_{\sig NN} \sig \bar NN$, giving the equation of motion 
    \begin{equation}
( \partial^2 + m_\sig^2) \sig = g_{\sig NN} \bar NN \;.
 \end{equation} 
  Going to the non-relativistic limit and assuming the nucleons to be a point source 
  $\bar NN(x)  \to  N^\dagger N \de^{(d-1)}(\veci{x})$, 
  as appropriate for the  far-field, 
 one gets  with $N^\dagger N =1$ 
 \begin{equation}
 \label{eq:sigeom}
 (-\veci{\partial}^2 + m_\sig^2)\sig = g_{\sig NN}\de^{(\dbar)}(\veci{x}) \;, 
 \end{equation}
 which is solved by the standard Green's function $\sig(r)  =  g_{\sig NN} G_{d-1}( m_\sig r )$:
   \begin{equation}
   (-\veci{\partial}^2 + m^2)G_n(r) = \de^{(n)}({\veci{x}})  \quad \Rightarrow \quad
   G_n(r) = \frac{1}{(2\pi)^{n/2}}
\left(\frac{m}{r}\right)^{\frac{n}{2}-1}
K_{\frac{n}{2}-1}(mr)  \;.
   \end{equation}
   
  Temporarily focusing back on $d=4$,   the $\sig$ far-field   reduces to
\begin{equation}
\label{eq:classical}
\sig(r)|_{r \to \infty}  =  g_{\sig NN} \frac{e^{- m_\sig  r}}{4 \pi r}  \;,
\end{equation}
 connecting to \eqref{eq:B} and changing to our units \eqref{eq:a} one gets
\begin{equation}
\label{eq:sol}
\hXX(r)|_{r \to \infty} = 1 -  \frac{ e \,  g_{\sig NN  } } {4 \pi \rr}    \frac{e^{- \hat{m}_\sig  r}} { r} \;,
\end{equation}
such that $A$ in \eqref{eq:B} becomes 
\begin{equation}
\label{eq:B2} 
A = \frac{  e\, g_{\sig NN  } } {4 \pi \rr} = \frac{m_N}{4 \pi \rr^2} \left[ \frac{e}{F_\pi}\right] = \frac{M_0}{ 4 \pi  \rr^2}   \;.
\end{equation}
In the second equality, we have assumed   the dilaton Goldberger-Treiman relation and in the last one 
 expressed the
$m_N$ in terms of \eqref{eq:M0}, thereby 
 ignoring the quantisation part. 
 This last step is appropriate since the energy function is minimised without the quantisation part (see the discussion in \cite{Cebulla:2007ei}). 
  We will see that the numerical  simulation  reproduces this asymptotic behaviour precisely for $m_\pi$ towards zero. 

As a next step we will compute the residue $r^N$ \eqref{eq:rN} from the gravitational form factor, using 
the solution in \eqref{eq:sol}. To do this, we need an expression of the residue in terms of $T_{\mu\nu}(x)$ 
which is expressible in terms of the $f$- and $\hXX$-fields.   
This can be achieved by considering the form factor in the Breit-frame  ($q = (0 , \veci{q})\;, q^2 = - \veci{q}^2$) 
where  one has  (see  \EQ 17b \cite{Polyakov:2018zvc})  
\begin{equation}
\matel{N(p')}{T_{ij}}{N(p)} = 2 m_N E_N (q_i q_j + \de_{ij} q^2 ) \frac{D(q^2)}{4 m_N^2} \;.
\end{equation}
Contracting indices and using  $\vev{N(p')|N(p)} = (2\pi)^{d-1} 2 E_N \de^{d-1}(\veci{p} - \veci{p}')$ 
(relativistic state normalisation) and the identification $(2\pi)^{d-1} \de^{d-1}(0) \leftrightarrow \int d^{d-1} x = V$ one gets the 
desired formula
\begin{equation}
\label{eq:rN2}
r^N =  \lim_{q^2 \to 0}  q^2 D(q^2)  = \frac{4 m_N}{d-2} \int d^{\dbar} x \, T_{ii}(\veci{x}) \;.
\end{equation}
This equation looks puzzling at first sight, as it is telling us that the theory has a (massless) dilaton unless 
the  integral on the right-hand side vanishes.  Hence, there must be such a condition 
in the generic case. This relation is known as  the von Laue condition (see the review \cite{Polyakov:2018zvc})
\begin{equation}
\label{eq:vL}
\vev{p}_V \equiv \int d^{\dbar}  x\, p(\veci{x})  = \frac{1}{\dbar}  \int d^{\dbar} x \, T_{ii}(\veci{x}) =0 \;,
\end{equation}
which states that the  pressure  integrated over the volume $V$ must vanish.\footnote{In this form the theorem 
is valid for spin $0$ and spin $1/2$ cf.  \cite{Polyakov:2018zvc}.}
 The derivation uses the conservation of the energy-momentum 
tensor and integration by parts. In the dilaton case, it is the latter assumption which breaks down.

The key quantity must be the improvement term as it is the case in the momentum space derivation 
in leading-order perturbation theory \cite{Zwicky:2023fay}. The latter is modified in $d$ dimensions to 
\begin{equation}
\label{eq:imp}
\Lag_d = - \frac{\xi_d}{2} R  F_\sig^2 \hat{\chi}^2  \quad  \to \quad   T_{\mu\nu} = \xi_d( \eta_{\mu\nu} \partial^2- \partial_\mu \partial_\nu) F_\sig^2 \hat{ \chi}^2 \;,
\end{equation}
where $\chi \equiv F_\sig/\dphi \hat{\chi}^{\dphi} $  is the naturally normalised scalar with dimension $ \dphi \equiv (d-2)/2$, 
and $\xi_d = \frac{\dphi}{2 (d-1)}$ is the improvement coefficient.
 Then we find,
\begin{equation}
\label{eq:Poisson}
T_{ii}(\veci{r}) =   2  \xi_d \dphi  \, \veci{\partial}^2   \chi^{2} = 4 \xi_d F_\sig g_{\sig NN}  \de^{(\dbar)}(\veci{x})=  4 \xi_d m_N  \de^{(\dbar)}(\veci{x}) \;,
\end{equation}
upon using  $ \chi^{2} =  \frac{F_\sig^2}{\dphi^2}(1 - 2 \dphi \sig/F_\sig + \dots)$,  
 the equation of motion \eqref{eq:sigeom} for $m_\sig=0$ and the dilaton Goldberger-Treiman relation \eqref{eq:dGT}.
 This expression modifies the von Laue condition $\vev{p}_V^{\text{von Laue}}$ to a dilaton version
 \begin{equation}
 \label{eq:vL2}
 \vev{p}_V^{\text{dilaton}} = \frac{4 \, \xi_d}{ d-1} \, m_N  \to \frac{2}{9} m_N \;.
 \end{equation}
Finally with \eqref{eq:rN2} we find the  residue \eqref{eq:rN}
\begin{equation}
r^N  = \frac{4 m_N}{d-2} \int d^{\dbar} x \, T_{ii}(\veci{x}) = 
  \frac{4}{d-1} m_N^2 \;,
\end{equation}
ensuring    IR-conformality $\matel{N(p)}{\TEMT}{N(p)} = 0$ for $m_N \neq 0$. 

Formally, the situation is analogous to classical electrodynamics, where $\hXX$ plays the role of the scalar potential in the Poisson equation for a point charge distribution. In our case, the role of the point charge is taken on 
by  the  nucleon at far distance.

\section{Observables, results and fits}
\label{sec:pred}

We consider nine observables in the fit, including  the previously discussed $F_\pi$, $m_N$, and $m_\Delta-m_N$, together with one magnetic moment and five radii obtained from the electromagnetic and weak form factors, see \APP\ref{app:EM} and \ref{app:weak}.
The experimental input is given in  the respective tables.

\subsection{Static nucleon observables}

We briefly summarise all observables in terms of seven integrals, denoted by $\{M_0, \rsES, \rsMS, \allowbreak \rsET, I_0, g_A^0, \rsA\}$, which are defined in \APP\ref{app:int}, \EQ\eqref{eq:int}.

Before discussing the nucleon mass, an important issue concerning the large-$N_c$ corrections from rigid-rotator quantisation needs to be addressed. 
Since, as previously stated,  the soliton profile is obtained by minimising the Lagrangian without the rigid-rotator contribution, adding the latter to the nucleon mass violates the von Laue condition. See Appendix C of \cite{Cebulla:2007ei} for an excellent discussion of this issue.\footnote{This should be distinguished from our special case of a massless dilaton, which leads to a well-controlled modification of the relation in \eqref{eq:vL2}.}
Following the consistent prescription advocated in \cite{Cebulla:2007ei}, we retain only the leading-order large-$N_c$ expression. This amounts to dropping the $\ORD(1/N_c^2)$ rigid-rotator term in \eqref{eq:mNDe} and considering the nucleon--$\De$ mass splitting rather than the $\De$ mass itself\begin{equation}
\label{eq:mNDe2}
m_N|_{\text{fit}}  = M_0\left[\frac{F_\pi}{e}\right]  , \qquad
\De m \equiv  m_\Delta - m_N  =   \frac{6}{I_0}  \left[e^3 F_\pi\right] \;.
\end{equation}

From the electromagnetic form factors we may define 
the four electromagnetic radii defined from the slopes of the electromagnetic form factors, 
separating  into  isospin $I = 0$ and $I=1$,  
\begin{alignat}{4}
 &\vev{r^2_E}_{I=0} &\; =  \;&   \rsES \left[ \frac{1}{e^2 F_\pi^2} \right]  \;, \quad 
&   &\vev{r^2_M}_{I=0} &\; =  \;&   \rsMS \left[ \frac{1}{e^2 F_\pi^2} \right]  \;,
  \nonumber   \\[0.1cm]
 &\vev{r^2_E}_{I=1} &\; =  \;&   \rsET \left[ \frac{1}{e^2 F_\pi^2} \right]  \;, \quad 
&   &\vev{r^2_M}_{I=1} &\; =  \;&   \rsMT \left[ \frac{1}{e^2 F_\pi^2} \right] \;.
\end{alignat}
and the two  magnetic moments  
\begin{alignat}{4}
\label{eq:mu}
 & \mu_{I=0}   &\; =  \;&  \frac{   \vev{r^2_E}_{I=0}}{6 \Isky } &\; =  \;&  \frac{ m_N  \vev{r^2_E}_{I=0}}{3 \Isky } \, [ \mu_N]  
 &\; =  \;& \frac{m_N \rsES}{3 I_0} \left[ \frac{e}{F_\pi} \right]  \, [ \mu_N]   \;,  \nonumber   \\[0.1cm]
&  \mu_{I=1}   &\; =  \;&  \frac{1}{3} \Isky  &\; =  \;&  \frac{2}{3} m_N \Isky \,  [ \mu_N ]  &\; =  \;&  \frac{2}{3} m_N I_0 \left[\frac{1}{e^3 F_\pi}
\right]  \,  [ \mu_N ]   \;.
\end{alignat}
Note that 
in the Skyrme model the $I=1$ radii are degenerate $ \rsET  =  \rsMT $, in slight tension with experiment 
as can be seen from \TAB\ref{tab:ex}.
A useful conversion factor is 
\begin{equation}
 \left[ \frac{1}{e^2 F_\pi^2} \right] =  \left[ \frac{\small \GeV^2}{ e^2 F_\pi^2} \right] 
(0.1973 \fm)^2 \approx (0.42 \fm)^2 \;,
\end{equation}
where  
$\left[ \frac{\small \GeV^2}{ e^2 F_\pi^2} \right]  \approx (2.15)^2$ for   reference values of $F_\pi = 93 \MeV$ and $e = 5$. 

From the weak form factor  $g_A(q^2)$, we get the nucleon axial charge  
and the nucleon  axial radius \eqref{eq:axial}
\begin{equation}
\label{eq:gA2}
g_A =  g^0_A \left[  \frac{1}{e^2} \right] \;, \qquad   \vev{r_A^2}   = 
\frac{\rsA}{g_A^0}\left[  \frac{1}{e^2 F_\pi^2} \right] \;.
\end{equation}

The asymptotic behaviour of   $B=  r^2  \exp(  \hat{m}_\pi r)  f(r)|_{r \to \infty}  $ \cite{Adkins:1983ya}, 
see the discussion in \APP\ref{app:con},  is given by
\begin{equation}
B = \frac{3}{8 \pi } \frac{ g_{\pi NN}}{ F_\pi m_N} [F_\pi^2 e^2] \;,
\end{equation} 
which upon using the pion Goldberger-Treiman relation $g_{\pi N N} F_\pi = g_A  m_N$ 
and the $g_A$-expression above assumes the form 
\begin{equation}
\label{eq:GTB}
B = \frac{3}{8 \pi} g_A^0 \;,
\end{equation}
which is exactly satisfied in the zero-pion-mass limit.

Comparing the observables obtained in the Skyrme model to real-world physical observables is not straightforward. 
The latter  are at fixed $N_c  = 3$, whereas the Skyrme model misses the $1/N_c$-suppressed pion loops. 
Perfect agreement should therefore not be expected, and a corresponding large-$N_c$ uncertainty must be assigned. For definiteness, we take this uncertainty to be $10\%$ for all observables.

Since we intend to fit the three parameters $F_\pi$, $e$ and $\rr$, we must identify observables that receive corrections well beyond this expectation.
For the quantities $\mu_{0,1}$, this has already been done 
in the pioneering paper \cite{Adkins:1983hy}, but in recent years more and more is known on 
large $N_c$ from lattice QCD (see \REFs\cite{Lucini:2012gg,Hernandez:2020tbc} for reviews). For other observables, this is assessed in 
  \APP\ref{app:largeNc} 
where it is found that $g_A$ and  \eqref{eq:GTB} have large corrections in $1/N_c$ and are therefore not suitable for a
fit since they would completely bias the fit-procedure due to the large corrections.

\begin{table}[ht]	
%\begin{sidewaystable}
\renewcommand{\arraystretch}{1.1}
\centering

\begin{tabular}{l | r ||  >{\columncolor{blue!10}} r | r ||  r | r}
$\veci{\text{Obs}} \quad (m_\pi \neq 0)$ & Exp & Sky-dil & $\veci{\De}_{\text{Obs}}[\%]$ & Sky & $\veci{\De}_{\text{Obs}}[\%]$ \\[0.05cm] \hline \hline
$F_\pi[\MeV]$       & $92.22(14)$  & $84.7$   & $-8$  & $70.4$   & $-24$ \\
$m_N[\MeV]$         & $939.6$      & $961.9$ & $2$   & $1082.5$ & $15$ \\
$(m_\Delta-m_N)[\MeV]$    & $292.4^\dagger$ & $288.8$ & $-1$ & $315.7$ & $8$ \\
$r_{E,I=0}[\fm]$    & $0.7690(13)$ & $0.68$    & $-11$  & $0.57$    & $-26$ \\
$r_{M,I=0}[\fm]$    & $0.82(9)$    & $0.86$    & $5$  & $0.82$    & $0$ \\
$r_{E,I=1}[\fm]$    & $0.9068(11)$ & $0.91$    & $1$   & $1.05$    & $16$ \\
$r_{M,I=1}[\fm]$    & $0.856(16)$  & $0.91$    & $7 $  & $1.04$    & $23$ \\
$\mu_{I=0}\,[\mu_N]$      & $0.8798$     & $0. 74$    & $-15$ & $0.62$    & $-29$ \\
$r_{A}[\fm]$      & $0.559(33)$     & $0.51$    & $-9$ & $0.34$    & $-40$ \\
%$\mu_1[\mu_N]$      & $4.7059$     & $4.23$    & $-10.1$ & $2.79$    & $-40.7$ \\
%$g_A$               & $1.2754(13)$ & $1.08$    & $-15.4$ & $0.95$    & $-25.1$ \\
%$g_{\pi NN}$        & $13.13(3)$   & $12.86$   & $-1.8$  & $14.88$   & $13.6$ \\
 \hline \hline
$\chi^2/\dof$        &              & $0.91$    &         & $6.08$    & \\ 
\end{tabular}

\caption{\small 
The fit values are $e_{\text{Sky-dil}} =  4.37$ and $\rr = 1$ for the Skyrme-dilaton, and  
$e_{\text{Sky}} = 4.86$ for the Skyrme.
  All values are from the PDG
\cite{ParticleDataGroup:2026mpi}
except the FLAG average
\cite{FlavourLatticeAveragingGroupFLAG:2024oxs}
of the pion decay constant and the axial radius is taken from  \cite{Alvarado:2026kny}.
%and $g_{\pi NN}$ from the $\pi$-$N$ scattering analysis \cite{Pavan:1999cr}.
The error on the nucleon mass and $\mu_{I=0}$ are negligible.
$^\dagger$Whereas the $\De$ is unstable with a pole of
$\sqrt{s_\De} = m_\De-\frac{i}{2}\Ga_\De
\approx (1232-\frac{i}{2}117)\MeV$,  we do not assign an
extra uncertainty as this will be covered by the generic $10\%$
large-$N_c$ model uncertainty assigned.
% and the
%smallness of the $g_A$ uncertainty derives from abundant
%$\beta$-decay measurements.
}
\label{tab:ex}

%\end{sidewaystable}
\end{table}

\begin{table}
\renewcommand{\arraystretch}{1.1}
\begin{centering}
\begin{tabular}{l | r  ||   >{\columncolor{blue!10}} r | r || r | r}
$\veci{\text{Obs}} \quad (m_\pi =0) $& Exp & Sky-dil & $\veci{\De}_{\text{Obs}}[\%]$ & Sky & $\veci{\De}_{\text{Obs}}[\%]$ 
\\ \hline \hline
$F_\pi[\MeV]$       & $80.3(3)$    & $87.1$   & $8$   & $63.16$   & $-21$ \\
$m_N[\MeV]$         & $836.24$  & $824.3$  & $-1$   & $884.4$  & $6$ \\
$(m_\Delta-m_N)[\MeV]$     & $260.4$ & $268.4$ & $3$   & $254.4$ & $-2$ \\
$r_{E,I=0}[\fm]$    &  $0.7690(13)$     & $0.75$    & $-2$   & $0.63$    & $-17$ \\
$r_{M,I=0}[\fm]$    & $0.82(9)$    & $1.0$    & $23$   & $0.99$    & $20$ \\
$\mu_{I=0}\,[\mu_N]$      & $0.88$    & $0.72$    & $-18$ & $0.52$    & $-41$ \\
$r_{A}[\fm]$      & $0.559(33)$     & $-$    & $-$ & $0.48$    & $-15$ \\
%$\mu_1[\mu_N]$      & $4.71$    & $4.41$    & $-6.4$  & $4.79$    & $1.7$ \\
%$g_A$               & $1.27$    & $1.14$    & $-10.7$ & $0.76$    & $-40.4$ \\
%$g_{\pi NN}$        & $13.1$    & $12.57$   & $-4.1$  & $11.82$   & $-9.8$ \\ 
\hline \hline
$\chi^2/\dof$    &    &  2.18          &           &   5.69      &
\end{tabular}
\caption{\small  See  \TAB\ref{tab:ex}, and text for which observables have been corrected for $m_\pi =0$. 
The fit values are $e_{\text{Sky-dil}} =  4.88$ and $\rr = 1.6$ for the Skyrme-dilaton, and  
$e_{\text{Sky}} = 5.69$ for the Skyrme.}
	\label{tab:ex0}
	\end{centering}
\end{table}

\subsection{Numerical solutions}

\begin{figure}[h!]
\centering 
\includegraphics[width=0.7\linewidth]{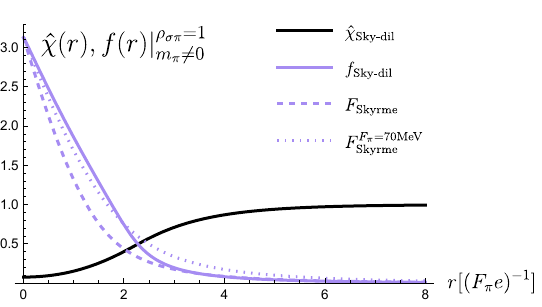}
%\vspace{-0.5cm}
	\caption{\small Plot showing the dilaton function $\hXX(r)$ (solid black) as a function of $r$ in our units of $1/(e F_\pi )= 1/( 5 \cdot 93\MeV)  \approx 0.424\fm$ (such that 2 units correspond to the proton charge radius). We also show the Skyrme-dilaton function $f(r)$ (solid purple) versus 
	the Skyrme function $F(r)$ (purple dashed) for physical $F_\pi$ and (purple dotted) for $F_\pi = 70 \MeV$ close
	to the best fit-value in that model. 
	The value with $\rr=1$ has been chosen for the Skyrme-dilaton  model and $m_\pi$ is set to its physical value. }
	\label{fig:comp}
\end{figure}

The numerical solutions are found by minimising the energy functional, which is equivalent 
to solving the equation of motion for the static case.  The strategy for finding the solutions is described 
in \APP\ref{app:com} alongside  plots with 
 the Skyrme $f(r)$ and the dilaton $\hXX(r)$ profile functions   in \FIG\ref{fig:fX} for various values of $\rr$. 
 For illustrative purposes, a basic plot for $\rr = 1$, comparing the Skyrme-dilaton and the Skyrme model, is shown in \FIG\ref{fig:comp}.
  There are two non-trivial 
test for the solutions. 

%% OLD TABLE - DO NOT USE
%% 
%% \begin{table}[ht]
%% \centering
%% \renewcommand{\arraystretch}{1.2}
%% \begin{tabular}{l | rrrr rrrrr}
%% $\rr$ & $2.0$ & $1.8$ & $1.6$ & $1.4$ & $1.2$ & $1.0$ & $0.8$ & $0.6$ & $0.4$ \\ \hline
%% $m_\sig[m_\pi]   $ & $1.00$ & $1.11$ & $1.25$ & $1.42$ & $1.67$ & $2.00$ & $2.50$ & $3.33$ & $5.00$  \\ \hline
%% $A_{m_\pi \neq 0} $ & $1.2$ & $1.5$ & $1.9$ & $2.6$ & $3.7$ & $5.7$ & $10.8$ & $\approx 28.8$ & $\approx 170$ \\ \hline
%% $A_{m_\pi = 0}     $   & $1.17$ & $1.41$ & $1.72$ & $2.15$ & $2.78$ & $3.78$ & $5.5$ & $9.0$ & $18.31$ \\ \hline
%%  $\hat{A}$     & $0.98$ & $0.94$ & $0.91$ & $0.83$ & $0.75$ & $0.66$ & $0.51$ & $0.31$ & $0.11$
%% \end{tabular}
%% \caption{ Above $\hat{A} \equiv \frac{A_{m_\pi = 0}}{A_{m_\pi \neq 0} } $.
%% The quantity $A_{m_\pi = 0} $ is the one given in \eqref{eq:sol} as extracted from the asymptotic form 
%% which equals the prediction in \eqref{eq:B2} to the given numerical precision. For $m_\pi \neq 0$ the non-zero 
%% $m_\sig = 2 m_\pi/\rr$ \eqref{eq:msig}, leads to corrections in the dilaton Goldberger-Treiman relation \eqref{eq:GT2} as given in the last row. 
%% %Incidentally, the (pion) Goldberger-Treiman relation is satisfied to about $2\%$ accuracy in the real world 
%% %which is of the same order as for $m_\sig = m_\pi$ in our model and thus qualitatively credible.  
%%   }
%% \label{tab:A}
%% \end{table}

\begin{table}[ht]
\centering
\renewcommand{\arraystretch}{1.2}
\begin{tabular}{l | rrrr rr}
$\rr$ & $2.0$ & $1.8$ & $1.6$ & $1.4$ & $1.2$ & $1.0$ \\ \hline
$m_\sig[m_\pi]   $ & $1.00$ & $1.11$ & $1.25$ & $1.42$ & $1.67$ & $2.00$ \\ \hline
$A_{m_\pi \neq 0} $ & $1.2$ & $1.5$ & $1.9$ & $2.6$ & $3.7$ & $5.7$ \\ \hline
$A_{m_\pi = 0}     $   & $1.17$ & $1.41$ & $1.72$ & $2.15$ & $2.78$ & $3.78$ \\ \hline
 $\hat{A}$     & $0.98$ & $0.94$ & $0.91$ & $0.83$ & $0.75$ & $0.66$
\end{tabular}
\caption{ Above $\hat{A} \equiv \frac{A_{m_\pi = 0}}{A_{m_\pi \neq 0} } $.
The quantity $A_{m_\pi = 0} $ is the one given in \eqref{eq:sol} as extracted from the asymptotic form 
which equals the prediction in \eqref{eq:B2} to the given numerical precision. For $m_\pi \neq 0$ the non-zero 
$m_\sig = 2 m_\pi/\rr$ \eqref{eq:msig}, leads to corrections in the dilaton Goldberger-Treiman relation \eqref{eq:dGT} as given in the last row. 
%Incidentally, the (pion) Goldberger-Treiman relation is satisfied to about $2\%$ accuracy in the real world 
%which is of the same order as for $m_\sig = m_\pi$ in our model and thus qualitatively credible.  
  }
\label{tab:A}
\end{table}

\begin{itemize}
\item [a)] The Skyrme virial theorem \eqref{eq:power}, is satisfied within numerical precision.
\item [b)] The asymptotic form \eqref{eq:sol}, in particular $A$ given in \eqref{eq:B2}. 
For $m_\pi =0 $ (which here implies $m_\sig =0$) 
where conformal symmetry appears unbroken,  disregarding potential changes due to quantum corrections, 
we expect $A$ to be matched exactly and this is indeed the case to within numerical accuracy.  
For $m_\pi \neq 0$ the dilaton Goldberger-Treiman relation ceases to be exact, in particular 
because $m_\sig = 2 m_\pi /\rr \neq 0$ \eqref{eq:msig} and there are corrections which are modest 
for low $\sig$-masses that is $\rr$ large, as can be inferred from \TAB\ref{tab:A}.
\end{itemize}

\subsection{$\chi^2$-fits to nucleon observables}

%:
In the Skyrme-dilaton model, there are four parameters, $(F_\pi,e,\rr,\hat{m}_\pi)$, as introduced in Eq.~\eqref{eq:four}. We fix
\begin{equation}
\hat{m}_\pi = \frac{m_\pi}{eF_\pi} \;, \qquad
\rr = \frac{F_\sig}{F_\pi} \;,
\end{equation}
and fit $F_\pi$ and $e$ for a range of values of $\rr$, while imposing the physical pion mass, $m_\pi=139.6 \,\MeV$ \cite{ParticleDataGroup:2026mpi}. This parametrisation is convenient because the equations of motion must be solved separately for each fixed pair $(\rr,\hat{m}_\pi)$. Since $\rr$ is empirically expected to be of 
order unity (\cite{Zwicky:2023krx} and references therein), we have scanned the range $0.4 < \rr < 2$.

The standard $\chi^2$ is defined by 
\begin{equation}
\chi^2 =   ( \veci{O}_{\,\text{Skyrme}}- \veci{O}_{\,\exp})^T C^{-1}  (\veci{O}_{\,\text{Skyrme}}- \veci{O}_{\,\exp}) \;,
\end{equation}
where $\veci{O}$ is the array of observables given in \TAB\ref{tab:ex}.  The uncertainties entering the fit are encoded in the covariance matrix $C$.   For the $\De$-baryon which is unstable the-width-mass-ratio is 
$\Ga_{\De}/m_{\De} \approx 0.1$ ($\Ga_\De \approx 117\MeV$)   but since the width is large-$N_c$ suppressed
we do not assign an additional uncertainty to it.

In addition to the experimental uncertainties, we must account for the theoretical uncertainties associated with the large-$N_c$ approximation and the incomplete treatment of strange-quark mass effects (linked to  the small  $\sig$ mass 
as previously discussed). Since our strategy is to exclude observables with systematically large $1/N_c$ corrections, such as $\mu_{I=1}$, $g_A$ and the pion Goldberger--Treiman asymptotics, from the fit, we assign  a total $10\%$-systematic  uncertainty as we simply cannot expect results to be better than that. 
 This is in line with the long-standing rule of thumb for well-behaved large-$N_c$ observables in the literature \cite{Witten:1984cg}.
  %We further assign a $5\%$ theoretical uncertainty to account for the incomplete treatment of the strange quark mass. The size of this effect can be estimated from the difference between the $SU(3)$ and $SU(2)$ chiral limits of the pion decay constant,  ${F}_\pi|_{m_{uds} \to 0} = 80.3(6) \MeV $ and
%$F_\pi|_{m_{ud} \to 0} \approx 87.5\MeV$.

The experimental and theoretical uncertainties are then added in quadrature to construct the covariance matrix.
\begin{equation}
C =\text{diag}   ( (0.1\, O_{\exp})^2    + \De_{\exp}^2 ) \;.
\end{equation}

 \subsubsection{The Skyrme-dilaton model}

 For physical pion mass, with the procedure described above,  we get  
 \begin{equation}
 \label{eq:main} 
 \chi^2/\dof  \big|_{\text{SD},m_\pi \neq 0} =  0.91 \;, \qquad  F_\pi = 84 .7 \MeV \;, \qquad e = 4.37\;, \qquad \rr = 1  \;,
 \end{equation}
 a  $\chi^2/\dof$ close to one, resulting in a  $p$-value of $p= 0.49$, with  
  $F_\pi$  close to its physical value of $F_\pi \approx 92.2\MeV$. 
 The value of $e$ which is a parameter of the effective theory 
 is slightly lower than for the original Skyrme model.
   
It is also interesting to consider the massless-pion limit and assess whether the resulting physics remains credible.  
In the real world we expect most observables to be smooth in the chiral limit, exceptions are 
the $I=1$ electromagnetic charge radii which diverge  and therefore must  be discarded 
(and the axial radius has been found to be unstable and is discarded for the time being).  For the remaining observables 
$m_\pi$ (and in particular $m_\sig \to 0 $) therefore constitute valuable testing ground. 
For the uncertainty analysis, it would be desirable to know all observables in the $m_\pi \to 0$ limit. In particular, since the strange-quark mass is mainly modelled through $m_\sig \neq 0$, which nevertheless vanishes as $m_\pi \to 0$, we must consider the $m_{uds} \to 0$ limit since the strange quark is expected to contribute  significantly  to the $\sig$-mass in the dilaton case.
The pion decay constant and nucleon mass are among the observables that are well studied in this limit:  
 $F_\pi|_{m_{uds}\to 0} = 80.3(6) \MeV$ \cite{FlavourLatticeAveragingGroupFLAG:2021npn} (in \SEC 5.3.5), while $m_N$ is known to decrease by about $11\%$ (see the review \cite{Hoferichter:2025ubp}). We assume the same reduction for the $\De$-baryon. For the remaining three quantities  $\{ \vev{r^2}^{I=0}_{E,M},\mu_{I=0} \}$ we are not aware of chiral 
 estimates of the same reliability,
 but we may simply add $10\%$ uncertainty in quadrature to the $10\%$ which then raises to $\approx 14\%$ systematic uncertainty. 

 Taking these changes into account and dropping the two infinite radii, we get 
\begin{equation}
\label{eq:main2} 
 \chi^2/\dof\big|_{\text{SD},m_\pi = 0} = 2.18 \;, \qquad  F_\pi = 87.1 \MeV \;, \quad e = 4.88 \;, \quad \rr = 1.6 \;, 
\end{equation}
which has a $p$-value of $p = 0.088 $.  This is somewhat lower than in the massive pion case,
  as
one would expect on grounds of not adapting  the observables to the chiral limit. 
It is   remarkable that the model does not lose its ground when 
the pion and the $\sig$ are made massless.  Testing this matter was one of the main aims of this work.  
 
Let us return to the two observables, $g_A$ and $\mu_{I=1}$, which were omitted from the fit based on large-$N_c$ considerations. For the fitted values of $(F_\pi,e,\rr)$, they take the values
$g_A = (0.74,\,0.76)$ and
$\mu_{I=1} = (3.34,\,3.08)\,[\mu_N]$ 
in the massive and massless cases, respectively. While the values of $\mu_{I=1}$ are reasonably close to experiment and including it would not significantly alter the fit, the same is not true for $g_A$. 
 
Another aspect worthwhile emphasising  is that the asymptotic behaviour due to the classical analysis, predicted in \eqref{eq:B} and \eqref{eq:B2},
\begin{equation}
\hXX(r)|_{r \to \infty}  = 1 - \frac{M_0}{4 \pi \rr^2} \frac{e^{- \hat{m}_\sig r }}{r} \;,
\end{equation}
with $\hat{m}_\sig = 2\hat{m}_\pi/\rr$, holds 
exactly within numerical precision for the computer solution.  
%This is analogue of getting $g_{\pi NN}$ form 
%the asymptotic behaviour of $f(r)$   \cite{Adkins:1983ya}, as previously discussed. 

\subsubsection{The Skyrme model }

In order to put these results into perspective, it is useful to assess the 
standard Skyrme-model ($\hXX \to 1$) limit equally. We find
\begin{alignat}{3} 
&  \chi^2/\dof \big|_{\text{S},m_\pi  \neq 0}    &\; =  \;& 6.08  \qquad & & F_\pi = 70.4 \MeV \;, \qquad e =  4.86 \;,  \nonumber \\[0.1cm] 
&  \chi^2/\dof \big|_{\text{S},m_\pi = 0}    &\; =  \;& 5.69  \qquad  & & F_\pi = 63.2 \MeV \;, \qquad e = 5.69  \;,
\end{alignat}
 corresponding to a considerably larger $\chi^2/\dof$  (corresponding to $p \approx 10^{-5}- 10^{-7}$) than the ones 
 obtained in  the Skyrme-dilaton model.  
For the massive and massless 
case we have nine and seven observables respectively and the number of degrees of freedom are therefore seven and five respectively since we minimise over $F_\pi$ and $e$.   The values of $g_A = ( 0.71,0.67)$  and  
$\mu_{I=1} = (3.43,3.53)[\mu_N]$ are obtained in the massive and massless case, respectively.  
 The latter  are not too far off from experiment and 
 including them into the fit does, unlike $g_A$, not change the  overall picture of the fit considerably.   
The slightly different values than in the fits \cite{Adkins:1983ya,Adkins:1983hy} are due to the inclusion of the axial radius 
and the exclusion of $g_A$.

\section{Summary and conclusions}
\label{sec:conc}

We have extended the Skyrme model by incorporating the $\sigma/f_0(500)$ as a dilaton. The motivation for doing so is both phenomenological and theoretical. The $\sig$-meson is the lightest scalar resonance in QCD and therefore provides a natural additional low-energy degree of freedom, while its possible interpretation as a pseudo-Goldstone boson of spontaneously broken scale symmetry gives a symmetry principle governing its couplings. Since we have restricted our analysis to static  quantities, the broad width of the physical $\sigma$ can be represented by an effective stable scalar.

We solved the coupled equations for the pion and dilaton profiles and confronted the resulting model with nine static nucleon observables. The parameters were determined from a combined $\chi^2$-fit, with uncertainties enlarged to account for the expected limitations associated with the large-$N_c$ nature of the Skyrme model. 
For a physical pion mass and $\rho_{\sigma\pi}=F_\sigma/F_\pi=1 $,\footnote{We note that $\rr =1$, 
also supported by fits to gravitational form factors \cite{Stegeman:2025sca,Stegeman:2025tdl},  is precisely the value 
for which the $\sig$-meson can mimic the Yukawa mechanism and becomes a candidate for a composite Higgs boson, 
see \cite{Cata:2018wzl,Zwicky:2023krx}.} we obtain \eqref{eq:main}
\begin{equation}
\chi^2 /\dof|_{m_\pi \neq 0}  = 0.91 \;,
\qquad
F_\pi = 84.7 \MeV \;,
\qquad
e = 4.37 \;,
\end{equation}
which is the main result of this paper. 
The quality of the fit is therefore substantially better than for the minimal Skyrme model, for which we find $\chi^2/\mathrm{dof} \approx  6.1$. In particular, the fitted value of $F_\pi$ is considerably closer to its physical value 
which was the main challenge for the original Skyrme model, see however \APP\ref{app:largeNc} for a more nuanced assessment. 
The improvement indicates that the $\sig$ plays a significant phenomenological role and  at this level is consistent with
 coupling like a dilaton.

Besides investigating the real-world impact of the $\sig$ coupled as a dilaton on physical observables,
 one might wonder what happens in the more extreme case of a massless dilaton.   
 Is it in principle possible that the $\sig$-meson becomes massless in the chiral limit? 
 After adapting some observables to this limit, and discarding the ones that diverge, we obtained 
${\chi^2}/\dof |_{m_\pi = 0} \approx  2.2$ \eqref{eq:main2}
corresponding to a $p$-value of approximately $p \approx 0.09$. 
This is considerably lower than the fit at the physical pion mass. The result should, however, not be overinterpreted: in the present effective model the simultaneous limit $m_\pi\to0$ also drives $m_\sigma\to0$, while the treatment of the strange-quark mass and   corresponding observables is necessarily approximate. The limit demonstrates that it is not impossible 
that the $\sig$-meson becomes massless in the chiral limit. 

A theoretical result is the derivation of the dilaton far-field asymptotics \eqref{eq:sol}, linked 
to the  dilaton Goldberger-Treiman relation. 
 For $m_\pi = 0$ it is exact and verified by our numerical solution whereas for $m_\pi \neq 0$ 
 there are controlled deviations due to $m_\sig  = 2 m_\pi /\rr$ not being zero, see \TAB\ref{tab:A}.  Another curious aspect 
 is the amendment of the von Laue condition \eqref{eq:vL2} to the dilaton case. 

Finally, 
our results suggest that the dilaton provides a useful extension of the Skyrme model and, importantly, a concrete framework in which the connection between scale-symmetry breaking and the $\sig$-meson can be investigated. More broadly, our findings lend support to the possibility that the $\sig$-meson couples approximately as a dilaton, the pseudo-Goldstone boson of spontaneously broken scale symmetry, and are consistent with the picture of QCD approaching an infrared fixed point. How far the $\sig$-pole moves towards the infrared in the chiral limit remains an open question. This can be further explored by scrutinising the model against Euclidean form-factor data and by incorporating more realistic effects of the strange-quark mass. Further progress on the axial nucleon charge $g_A$ would also be desirable, either from reliable 
large-$N_c$ estimates or by incorporating such effects into the Skyrme-dilaton approach.

\paragraph{Acknowledgements:}
We are grateful to   Christoph Adam, Mohamed Anber, Claudio Bonanno, Cedric Lorc\'e, Valya Khoze, Donal O'Connell, 
the late  Mannque Rho, Bernd Schroers, Paul Sutcliffe and Tom Winyard for discussions.   RZ is supported by  STFC via the consolidated grants ST/T000600/1 and ST/X000494/1.

\paragraph{Note added:}  

While completing this work, we became aware of related work \cite{prep} in which the $\sig$-meson is coupled to the BPS Skyrmion \cite{Adam:2010ds}. This approach shares certain features with, but also differs in important respects from, the dilaton model considered here.

\appendix

%\section{Plots of the solutions}
%\label{app:plots}

\section{Connection of the $\sig$- to the $\pi$-asymptotics in ANW }
\label{app:con}

In the pioneering work \cite{Adkins:1983ya}, a   connection is made between the asymptotic behaviour of the pion field and the pion Goldberger–Treiman relation, analogous to the discussion presented in \SEC\ref{sec:GT} for the dilaton case. The presentation, however, differs somewhat, and we therefore consider it worthwhile to parallel the two 
derivations.

  For that purpose it is useful to introduce 
the auxiliary form factor 
\begin{equation}
\tilde D(q^2) \equiv  \frac{3 q^2}{4 m_N^2}  D(q^2) \;,
\end{equation}  such that $\tilde{D} \equiv  \tilde D(0)$ is the analogue of  
 $g_A = g_A(0)$. In the massless dilaton case we have 
$\tilde D = 1$ (we work in the $m_\pi=0$ limit in this section). Their presentation in \SEC 4 \cite{Adkins:1983ya} proceeds in three steps:
 \begin{enumerate}
 \item \emph{Zero-momentum form factor via  full solution:} from  \eqref{eq:gA2} and \eqref{eq:rN2} we get
 \begin{equation}
 \label{eq:full}
 g_A  = g_A^0(f) \left[ \frac{1}{e^2} \right] \quad \leftrightarrow \quad \tilde D = 2 m_N \int d^3x \,  T_{ii}(f,\hXX) \;.
 \end{equation}
 \item \emph{Asymptotic value from  numerical simulation:}
 we know that the far-field is given by
\begin{equation}
\label{eq:as}
f(r) =   \frac{B}{{r}^2}  \quad \leftrightarrow \quad  \hat{\chi}(r)= A' -  \frac{A}{r} \;,
\end{equation}
and that for conserved currents ($m_\pi =0$) 
only the asymptotic form will play a role for the zero-momentum-transfer form factor. 
Hence, we may take the asymptotic ansatz, insert it into \eqref{eq:full}, and obtain expressions for 
$g_A \propto B$ and $\tilde D \propto A$, respectively. We can then take $B$ and $A$ from the numerical solution 
to confirm the same value as for the zero-momentum form factor. 
  This is a non-trivial tests of the formalism and the numerics. 
 \item  \emph{Goldberger-Treiman relation:}   
 one may do so by solving for the asymptotic fields classically as done in \eqref{eq:sigeom}, and obtain 
 \begin{equation}
 f(r)  = \frac{3}{8 \pi} \frac{g_{\pi NN}}{F_\pi m_N  } \frac{1}{r^2} \quad \leftrightarrow \quad  
 \hat{\chi}(r)= A' -  \frac{1}{4 \pi} \frac{g_{\sig NN}}{ F_{\sig} r} \;,
 \end{equation}
 where here $r$ has mass dimension of minus one (not rescaled as in \eqref{eq:a}) and 
 $A'=1$ by choice see the discussion around \eqref{eq:rescale}).
 The next step is to relate $g_{\pi NN}$ and $g_{\sig NN}$ to $g_A$ and $\tilde D=1$ as follows:
 \begin{equation}
 g_A = \frac{g_{\pi NN} F_\pi}{m_N} \quad \leftrightarrow \quad     1    = 
 \frac{g_{\sig NN} F_\sig}{m_N} \;,
 \end{equation}
 which reveals  the pion and dilaton Goldberger-Treiman relation which guarantee that the chiral and conformal 
 Ward identities are satisfied. 
  We reemphasise that the $m_N$ entering 
 the equation is only the classical part of the nucleon mass since we minimise with respect to the classical part. 
 \end{enumerate} 
 At last it seems worthwhile to contrast the two asymptotic formulae for $m_\pi \neq 0$. In this case they
 become 
  \begin{equation}
  \label{eq:asy}
 f(r)  = \frac{3}{8 \pi}  \frac{g_{\pi NN} }{F_\pi m_N} \frac{e^{- {m}_\pi r }} {  r^2} (1+ {m}_\pi r)  \quad \leftrightarrow \quad  
 \hat{\chi}(r)= 1 -  \frac{1}{4 \pi} \frac{g_{\sig NN }\, e^{-{m}_\sig r }}{F_\sig r} \;.
 \end{equation}
 where the form for $f$ is given in \cite{Adkins:1983hy} and the second one has been derived earlier \eqref{eq:sol} 
 (in  units where $r$ is dimensionless though).

\section{Large $N_c$ considerations}
\label{app:largeNc}

As the Skyrme model operates in the large $N_c$ limit it is important to assess which observables are reasonably accessible in such a limit. It is often said that the large $N_c$ limit works to $10\%$ which we take as a guideline 
for the discussion below and the uncertainty estimate.

A point raised in  \cite{Adkins:1983ya} is that  there are more observables than integrals, 
here \eqref{eq:int},  and this means that not all observables are independent within the model.  
In  \cite{Adkins:1983ya}  relations are given for the magnetic moments $\mu_{I=0,1}$. 
Whereas the one for $\mu_{I=0}$ is well-satisfied in nature, the one for $\mu_{I=1}$ is not. 
Hence, no solution can improve this matter, but rather one has to change the model more substantially 
by adding vector mesons for example as pointed out in    \cite{Adkins:1983ya}.
We therefore drop $\mu_{I=1}$ from the fit and quote its value independently.

A particularly delicate quantity is the weak axial charge, whose sensitivity has been discussed from various perspectives \cite{Meier:1996ng}. For example, in the naive quark model one finds $g_A \propto (1+2/N_c)$, suggesting potentially sizeable corrections unless significant cancellations occur  coming from other contributions. We will therefore not consider it 
for the fit as otherwise it will completely bias the resulting parameters.  Since we will find that $m_N$ and $F_\pi$ have similar large $N_c$ corrections, the Goldberger-Treiman relation  then implies that $g_{\pi NN}$ has corrections of the same order as $g_A$ and we can therefore not expect it to be well-behaved either. 

More promising observables are the radii, which are determined by the derivative of the form factor normalised to its value at zero momentum transfer. 
 The large $N_c$ emergent spin-flavour symmetry \cite{Gervais:1983wq,Dashen:1994qi,Manohar:1998xv}, 
 related to $m_\De - m_N = \ORD(1/N_c)$, 
 assures the electric and magnetic form factors $F$ have relative corrections 
 of  
 \begin{equation}
 F(q^2) = c( A(q^2) + B(q^2) \frac{J(J+1)}{N_c^2}) \;,
 \end{equation}
 such that   $F'(0)/F(0) = \ORD(1/N_c^2)$ 
 and  are therefore in the class of observables that are well-behaved in the large-$N_c$ limit. 
 This includes the axial radius.

 While the integrals in \eqref{eq:int} are, of course, independent of $N_c$, our parameters $F_\pi$ and $e$ are not. Their $N_c$-dependence must therefore also be taken into account, as it may distort the picture. A lot of knowledge has been accumulated
 in the last decade and it is found that many observables have rather sizeable large-$N_c$ corrections.   
 For example, using that $F_\pi = \ORD(\sqrt{N_c})$, one may define the appropriate quantity 
 \begin{equation}
  \tilde{F}_\pi \equiv  \frac{\sqrt{3} F_\pi}{\sqrt{N_c} F_\pi|_{N_c \to \infty}} = 
  1- \frac{0.92(14)}{N_c} + \ORD(1/N_c^2)  \approx  0.7 + \ORD(1/N_c^2)  \;,
 \end{equation}
 it is found that the pion decay constant is reduced by about $30\%$ in the linear fit \cite{Perez:2020vbn,Bonanno:2025hzr} which makes use of finite $N_c$ lattice data \cite{DeGrand:2023hzz}.  
 At first thought this calls into the question one of the main results here namely that $F_\pi$ is fitted 
 to be close to its physical value. In addressing this question we must also address other observables such as the 
 nucleon mass, since $m_N \propto F_\pi/e$. Defining $\tilde{m}_N$ similarly, taking into account that $m_N = 
 \ORD(N_c)$, one infers from \FIG 12 in \cite{Hernandez:2020tbc} (based on \cite{DeGrand:2013nna,DeGrand:2016pur}) 
 that  the nucleon mass also reduces considerably $\tilde{m}_N \approx  1- 1.14/N_c = 0.62.$\footnote{Even 
 though  this result is for a heavy pion mass $m_\pi /m_\rho \approx 0.55$, its size can be expected to be indicative 
 for the physical pion case:  $m_\pi /m_\rho \approx 0.18$.}
 If our result is to be meaningful, the change in $e$ has to be moderate. 

We are not aware of a direct study of the large-$N_c$ behaviour of $e$, and therefore take a detour by employing two phenomenologically successful relations: the KSRF relation, $g_{\rho\pi\pi}=m_\rho/(\sqrt{2}F_\pi)$, and the hidden local symmetry relation, $e=g_{\rho\pi\pi}$. Together, these give $e=\frac{m_\rho}{\sqrt{2}F_\pi}\approx 5.9$
which is reasonably close to the values obtained in Skyrme-model studies.\footnote{In reality it is more subtle since one needs to add pion loops to give meaning to $e$ which is thought to be matched at $\mu=m_\rho$ in the resonance 
saturation approach.} For our purposes, the main advantage of this connection is that it allows us to infer the large-$N_c$ behaviour of $e$ from that of $m_\rho$. Combining $m_\rho/\sqrt{\sigma}\approx 1.78$ with the large-$N_c$ correction of the string tension, $\tilde{\sigma}\approx 1-0.13/N_c=0.96$ \cite{DeGrand:2016pur}, which exhibits only a mild $N_c$ dependence, we obtain $\tilde{m}_\rho=0.85(5)$. This finally yields $
\tilde{e}=\frac{\tilde{m}_\rho}{\tilde{F}_\pi}\approx 1.2(1)$ 
which we use as a reference value.\footnote{The moderate large-$N_c$ dependence is consistent with the recent 
   S-matrix bootstrap  results in \cite{Cordoba:2026wsy}.}
 
 This provides further consistency checks, in particular as to whether imposing the physical values of $m_N$ and $F_\pi$ is meaningful within the Skyrme model. For this to be the case, the following ratio should be close to unity:
$\frac{\tilde F_\pi}{\tilde e\,\tilde m_N}\approx 0.96(10)$, 
which is indeed satisfied. Interestingly, this is also the combination that ensures that $\mu_{I=0}$ remains a well-behaved observable, as can be seen from \eqref{eq:mu}.
The mass splitting scales as $m_\Delta-m_N\propto e^3 F_\pi$, and the corresponding quantity, $\tilde e^3\tilde F_\pi=1.2(3)$, is not   small. Nevertheless, since we assign an additional uncertainty to this quantity to account for the finite $\Delta$ width, we retain it in our analysis.
Finally, all radii, being dimensionful quantities, scale as $\sqrt{\langle r^2\rangle}\propto 1/(eF_\pi)$. However, since $\tilde e\tilde F_\pi\approx\tilde m_\rho=0.85(5)$, as found above, the corresponding corrections remain within a tolerable range.

 In conclusion, we may keep all quantities but $\mu_{I=1}$, $g_A$ and must be cautious with the asymptotic 
 condition due to the pion Goldberger-Treiman relation. 

\section{Definition of observables}
\label{app:obs}

For completeness, we define the observables included in our fit, which are derived from the electromagnetic, weak and gravitational form factors in \SECs \ref{app:EM}, \ref{app:weak} and \ref{app:grav}, respectively.

\subsection{Electromagnetic form factors}
\label{app:EM}

The electromagnetic Dirac and Pauli form factors are defined by ($q=p'-p, \; N= p,n$) 
\begin{equation}
\label{eq:PD}
\langle N(p')|J^\mu_{ em}|N(p)\rangle
=
\bar u(p')\left[
\gamma^\mu F^N_1(q^2)
+i\frac{\sigma^{\mu\nu}q_\nu}{2m_N}F^N_2(q^2)
\right]u(p) \;,
\end{equation}
 and  related to the Sachs form factors by 
\begin{equation}
G^N_E(q^2)=F^N_1(q^2)+\frac{q^2}{4m_N^2}F^N_2(q^2) \;,
\qquad
G^N_M(q^2)=F^N_1(q^2)+F^N_2(q^2) \;.
\end{equation}
Their values at zero momentum transfers and their slopes define  prominent nucleon observables. 
For example,  the proton and neutron  magnetic moments 
\begin{equation}
G^p_M(0) =   \mu_p [\mu_N] \;, \quad G^n_M(0) =  \mu_n [\mu_N] \;,
\end{equation}
where $[\mu_N] = e_\text{el}/(2m_N)$ is the nucleon magneton and $e_\text{el}$ is the electric charge  
(charges are trivial $G^p_E(0) = Q_p = 1$ and $G^n_E(0) = Q_n = 0$).
In the Skyrme model it is customary to define the electromagnetic form factors without electric charge 
and thus $e_\text{el}$ is dropped from $[\mu_N]$. Moreover, because of isospin symmetry of QCD (for degenerate up and down quark masses), it is convenient 
to define the isospin combinations 
\begin{equation}
\label{eq:muI}
\mu_{I=0} =  \mu_p +  \mu_n \;, \qquad \mu_{I=1} =  \mu_p -  \mu_n \;.
\end{equation}
The slopes define the four radii
\[
\langle r_E^2\rangle_N = \frac{6}{G^N_E(0)}\left.\frac{dG^N_E(q^2)}{dq^2}\right|_{q^2=0}\;,\qquad 
\langle r_M^2\rangle_N = \frac{6}{G^N _M(0)}\left.\frac{dG^N_M(q^2)}{dq^2}\right|_{q^2=0} \;,
\]
from where we may define the isospin combinations 
\[
\langle r_E^2\rangle_{I=0}=\langle r_E^2\rangle_p+\langle r_E^2\rangle_n \;,\qquad
\langle r_E^2\rangle_{I=1}=\langle r_E^2\rangle_p-\langle r_E^2\rangle_n \;.
\]
and 
\[
\langle r_M^2\rangle_{I=0}=\frac{\mu_p\langle r_M^2\rangle_p+\mu_n\langle r_M^2\rangle_n}{\mu_p+\mu_n}\;,\qquad
\langle r_M^2\rangle_{I=1}=\frac{\mu_p\langle r_M^2\rangle_p-\mu_n\langle r_M^2\rangle_n}{\mu_p-\mu_n} \;.
\]

\subsection{Weak form factors}
\label{app:weak}

The weak form factors are defined by
\begin{equation}
\label{eq:weak}
\Ga_\mu(q^2) \equiv  \matel{N^{\bar{a}}(p')}{A_\mu^b}{N^c(p)} =   
\eps^{\bar a bc }\bar u (p')(\ga_\mu \ga_5 \, g_A(q^2) +  \frac{1}{2m_N} \ga_5 q_\mu \, g_P(q^2)  )u(p) \;,
\end{equation}
for $A_\mu^{b} = \bar q \ga_\mu \ga_5 T^b q$ with $T^b$ an $SU(2)$ flavour matrix. 
The weak charge and the so-called axial radius\footnote{See \REF\cite{Chen:2024oxx} for clarifying that it is rather a current spin density as the apparent analogy to the electric case does not carry over due to 
the $\ga_5$ in the current. }
 are defined by 
\begin{equation}
\label{eq:axial}
g_A = g_A(0) \;, \qquad   \vev{r_A^2} \equiv -  \frac{6}{g_A} \frac{d}{d q^2} |_{q=0} g_A(q^2) \;.
\end{equation}

 Since we are to use the Goldberger-Treiman relation 
to compute $g_{\pi NN}$, it seems worthwhile to sketch how it derives from \eqref{eq:weak}.
First, one writes 
\begin{equation}
g_A(q^2) = g_A + \ORD(q^2) \;, \qquad \tilde{g}_P(q^2) =  - \frac{q^2}{4 m_N^2} g_P(q^2)  = 
  g_{\pi N N}  \frac{F_\pi}{m_N}    \frac{ q^2 }{q^2- m_\pi^2} + \ORD(1) \;,
\end{equation}
with the pion pole appearing in the $g_P$ form factor, and $g_{\pi NN}$ can be defined from the effective Lagrangian 
\begin{equation}
\label{eq:gpiNN} 
\Lageff =  g_{\pi NN} \, \pi^a \bar N T^a i \ga_5 N \;.
\end{equation}
 Then,
one observes that for the chiral Ward identity  $q^\mu \Ga_\mu= \ORD(m_\pi^2)$ to be obeyed,
the pion Goldberger-Treiman relation \cite{Goldberger:1958tr}  
\begin{equation}
\label{eq:GT2}
g_{\pi N N} = g_A \frac{m_N}{F_\pi}  \;,
\end{equation}
must hold for massless pions.

\subsection{Gravitational  form factors}
\label{app:grav}

It is convenient to define $\bar u(p') \Theta^N_{\mu\nu}(q) u(p)  \equiv 2m_N  \matel{\ferm(p')}{T_{\mu\nu}}{\ferm(p)}$, 
from where the three form factors  follow
\begin{equation}
  \Theta^N_{\mu\nu}(q)  =   a_{\mu\nu} \,\gone^\ferm (q^2) + j_{\mu\nu} \,\gJ^\ferm(q^2) + 
d_{\mu\nu} \,\gtwo^\ferm (q^2) \;,
\end{equation}
 with ($\sig_{\mu \nu} = \tfrac{i}{2} [ \ga_\mu , \ga_\nu ]$, $\sig_{q\nu} = \sig_{\mu\nu} q^\mu$, $q= p' -p$ and ${\cal P} = \frac{1}{2}(p+p')$)
\begin{equation}
a_{\mu\nu} =2 {\cal P}_\mu {\cal P}_\nu \;, \quad d_{\mu\nu}  =\frac{1}{2} (q_\mu q_\nu-  q^2 \eta _{\mu\nu} ) \;, \quad
j_{\mu\nu} = 2 i \, {\cal P}_\mu \sig_{q \nu } + {\mu \leftrightarrow \nu } \;,
\end{equation}
Lorentz structures ensuring translational invariance  $q^\mu  \Theta^N_{\mu\nu} (q)= 0$.
The normalisation    $ A(0) =1 $ holds model-independently, since
 $T_{\mu\nu}$ is the associated Noether current with conserved momenta $P_\mu = \int d^3 x T_{\mu 0}$.
 Furthermore, for the nucleon one has $J^N(0) = \tfrac{1}{2}$, which reflects the absence of an anomalous gravitational magnetic moment, a consequence of the universality of
gravity. The form factors  $A^{N}$ and $J^N$ are analogues of the electromagnetic, Dirac and Pauli form factors, 
 given in \eqref{eq:PD}.
Contracting one gets
 \begin{equation}
  \label{eq:Ntrace}
  \Theta^N(q^2)    = 2 m_N^2 A^N(q^2) - \frac{q^2}{2} ( A^N(q^2)  - 2 J^N(q^2) + 3 D^N(q^2) )\;,
\end{equation}
and we see that $\Theta^N(0) = 2m_N^2$ $(\Theta = \Theta^{\rho}_{\phantom{\rho}\rho}$)  if all form factors are regular in the $q^2 \to 0$ limit 
but  $\Theta^N(0) = 0$  for a massless dilaton: $D^N(q^2) = \frac{4}{3} \frac{m_N^2}{q^2} + \ORD(1)$. 

\section{The seven basic integrals entering the observables}
\label{app:int}

For convenience, we reproduce all  integrals entering the observables (using   \eqref{eq:short})
\begin{alignat}{2}
\label{eq:int}
 & M_0(f,\hXX)    &\; =  \;&   2 \pi  \int_0^\infty  dr \, r^2  \left( \hat{\XX}^2 (   f'^2 + 2 \frac{s^2}{r^2}) +  \rr^2 \hat{\XX}^{'2} + 
 \frac{s^2}{r^2}(\frac{s^2}{r^2} + 2f'^2) + { \hat{m}_\pi^2  (1+\hXX^4  - 2\hXX^2 c    ) } \right)  \;, \nonumber   \\[0.1cm]
& I_0(f,\hXX)    &\; =  \;&   \frac{8\pi}{3}   \int_0^\infty dr \, r^2 s^2 \left(\hXX^2  +  (  f'^2 +  \frac{s^2}{r^2} ) \right)     \;, \nonumber   \\[0.1cm]
 & \rsES(f)     &\; =  \;&   \int_0^\infty dr \, r^2 \rho_B(r)    \;, \qquad \rho_B(r) = 4 \pi r^2 B_0(r) = - \frac{2}{\pi} f' s^2  \;,\nonumber   \\[0.1cm]
 & \rsMS(f)      &\; =  \;&  \int_0^\infty dr \, r^2 \rho_M(r) \;, \qquad   
 \rho_M(r) =  \frac{ r^2 \rho_B(r) }{ \int_0^\infty dr \, r^2 \rho_B(r)} \;,   \nonumber   \\[0.1cm]
 & \rsET(f,\hXX)       &\; =  \;&     \int_0^\infty dr \, r^2 {\rho}_{I=1}(r)  \;, \qquad  {\rho}_{I=1} =  \frac{\bar{\rho}_{I=1}}{\int dr \bar{\rho}_{I=1}} 
  \;,  \quad \bar{\rho}_{I=1} =  r^2 s^2\left(  \hXX^2 +  (f'^2+\frac{s^2}{r^2} ) \right) \;, \nonumber   \\[0.1cm]
 & g_A^0(f,\hXX)    &\; =  \;&  - \frac{4 \pi}{9}   \int dr  r^2 \left\{\begin{array}{rr}  \Om_-(r) & \quad m_\pi \neq 0 \\ 
  \frac{3}{2}  \Om_-(r) & \quad m_\pi =  0  \end{array} \right. \;, \nonumber   \\[0.1cm]
 & \rsA (f,\hXX)    &\; =  \;&   -\frac{4 \pi}{9} \int dr r^4  \left\{\begin{array}{rr}   (  \Om_-(r)  +  \frac{2}{5}    \Om_+(r)        )  & \quad m_\pi \neq 0 \\    \frac{3}{2} \Om_-(r)               & \quad m_\pi =  0  \end{array} \right. \;,
\end{alignat}
where the formulae of the two axial observables are sensitive to whether the pion is massive or not, 
as  explained  in  \SEC\ref{sec:axial} below. 
The remaining cases can be obtained from the literature by multiplying the kinetic contributions by a factor of $\hXX^2$, with some additional modifications for $M_0$, which are discussed in detail in the main text. When comparing with \cite{Adkins:1983ya}, one should also account for the different convention used for $F_\pi$, as explained in footnote \ref{foot:Fpi}.
 The expression for $\rho_B$ is unchanged since the dilaton does 
not couple to the baryon current and is easily computed from \eqref{eq:B2} and \eqref{eq:U0}; it is given 
below \EQ 18 in \cite{Adkins:1983ya}. The same applies to $\rho_{I=1}$, which has a dilaton addition in the 
term proportional to $F_\pi^2$ in \cite{Adkins:1983ya}.  The 
expression for $I_0$ agrees with \EQ 4.64 in \cite{Ma:2016npf} and can be shown to be equivalent to the 
expression given in \cite{Adkins:1983hy} upon taking the several conventions into account. The integral for 
$g_A^0$, after adding the $\hXX^2$-factor,  is equal to   \EQ 4.105 \cite{Ma:2016npf}, correcting an overall sign mistake, and  equivalent to \EQ 30 in \cite{Adkins:1983ya} after taking conventions into account.

\subsection{The axial observables}
\label{sec:axial}

The axial form factor $g_A(q^2)$ (and $g_P(q^2)$) have been derived in  \cite{Braaten:1986md} (\EQ 4.17). 
In the massless pion case, the axial Ward identity allows for simplifications and are at the origin of case distinctions 
in \eqref{eq:int}.
The generally valid result for the axial form factor reads
\begin{equation}
\label{eq:gAS}
g_A(q^2) =  \frac{1}{9 } \left( \vev{ \Om_+ }_2(q^2) -  \vev{ \Om_- }_0(q^2)  \right) \;. 
\end{equation}
where 
\begin{equation}
 \vev{ \Om_\pm}_n(q^2) = 4 \pi \int_0^\infty dr r^2 j_n(qr) \Om_\pm(r)  \;,
\end{equation}
with $\Om_+(r)   = \om_0 + \om_1(r)$ and $\Om_-(r)   = 2 \om_0 - \om_1(r)$ and
\begin{equation}
\om_0   =   
  \frac{s_2}{r} (\hat{\chi}^2 + f'^2 + \frac{s^2}{r^2})  \;,  \qquad   \om_1   =  - 2 f' ( \hat{\chi}^2 + 2 \frac{s^2}{r^2} )  \;,
\end{equation}
adapted to the dilaton case and taking into account their conventions.  
In the massless pion case, the axial current is conserved, $\partial \cdot A^a =0$ as a matrix 
element, implying relations amongst the axial and pseudoscalar form factor which amount to
\begin{equation}
\label{eq:magic}
\vev{ \Om_-}_0 + 2 \vev{ \Om_+}_2 \big|_{m_\pi=0} =0 \;,
\end{equation}
and this explains the two axial formulae in \eqref{eq:int}. We have checked that 
the formulae in  \eqref{eq:int} are numerically continuous in the $m_\pi \to 0$ limit.

\section{Plots and numerics}
\label{app:plots}

\subsection{Plots of Skyrme-dilaton solution for variable $\rr$}

\begin{centering}
\begin{figure}[h!]
\includegraphics[width=0.5\linewidth]{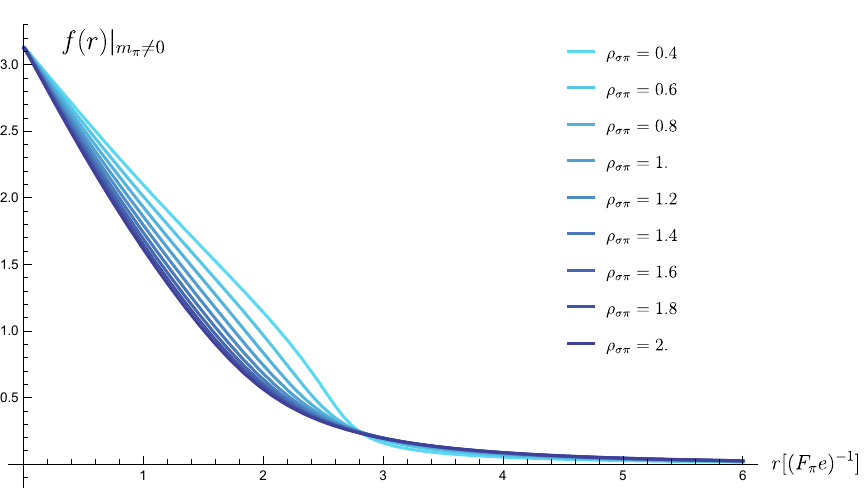} 
\includegraphics[width=0.5\linewidth]{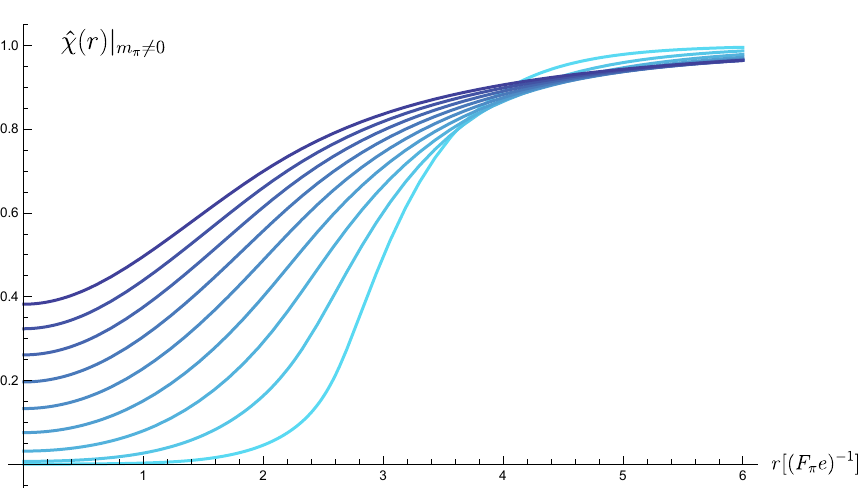} \\
\includegraphics[width=0.5\linewidth]{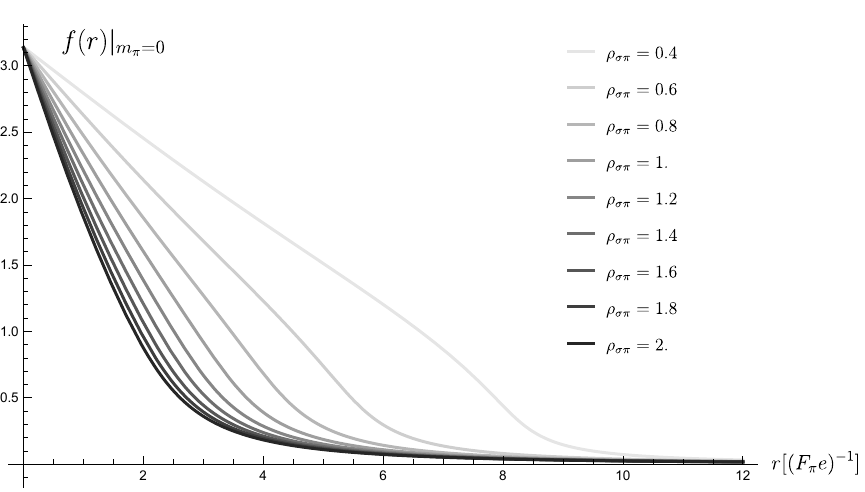} 
\includegraphics[width=0.5\linewidth]{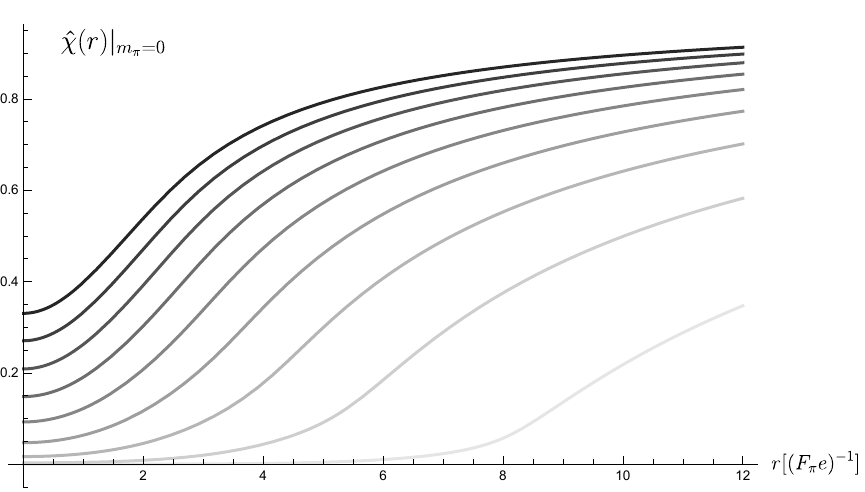} \\
%\vspace{-0.5cm}
	\caption{\small Plots of dilaton $\hXX(r)$ and Skyrme  $f(r)$ as a function of $r$ in our units of $1/(e F_\pi )$ 
	(top) physical pion mass  (bottom)  massless pion, for chosen values of $e F_\pi = 5 \cdot 93 \MeV$ implying 
	$\hat{m}_\pi = 46/155 \approx 0.297$. 
	 }
	\label{fig:fX}
\end{figure}
\end{centering}

Plots of the Skyrme-dilaton solution are given in \FIG\ref{fig:fX} for variable $\rr$ for the massless 
and the massive pion case. 

\subsection{Finding classical field configurations numerically}
\label{app:com}

While the equations of motion \EQs\eqref{eq:f} and \eqref{eq:chi} can be
solved numerically via a shooting approach, this is generally not a
promising strategy. If one were to start at $r=0$, radially integrate
outward and then try to use some optimization strategy to satisfy
boundary conditions at $r=\infty$, this would run into the same
problem that we would encounter when trying to find energy eigenstates
of the harmonic oscillator in such a way: there are different viable
options for large-$r$ behavior that correspond to different
solutions. For the harmonic oscillator, these differ in eigenstate
energy, while here, they differ in Baryon number. Being able to
``jump'' between such solutions tends to confuse whatever numerical
search strategy is employed (which may well include elements such as
penalty terms for trial solutions that depend on total energy) in
conjunction with a shooting method.

A more powerful -- but in terms of numerical algorithms still rather
basic -- approach is to instead use the Hamiltonian to work out the
energy of a classical configuration, discretise the problem by picking
sampling points at which the value of $f$ and $\hat{\chi}$ are used as the
degrees of freedom of the discretised approximation to the continuum
problem, then obtain $\partial_r f$ and $\partial_r \hat{\chi}$ via
algorithmic differentiation of some (e.g. cubic) spline interpolation
with continuous first derivatives and use that data alongside the
original interpolants to numerically estimate the total energy of the
candidate field configuration. Here, care has to be taken to employ an
integration scheme for energy density that is compatible with -- or
ideally directly derived from -- spline interpolation, for one can not
expect high-order Gauss-Kronrod quadrature to produce meaningful
results if interpolation order is truncated at some low polynomial
degree. It furthermore makes sense to transform the integration range
$r\in[0,\infty)$ to some finite interval by a transformation such as
  $q(r) = r/(1+r)$.

Numerical minimisation then yields useful approximations to the
desired ground state if one pins $f(0)=\pi$ and $f(\infty)=0$ in order
to enforce obtaining a $N_B=1$ solution, while also pinning
$\hat{\chi}(\infty)=1$. Even when not enforcing a fourth boundary condition,
(finite) energy minimisation produces the desired $\partial_r\hat{\chi}(0)=0$
solution.

The physical idea is that the value of the dilaton
field approaches some constant at infinity, and we can make that
constant equal to 1 by suitable redefinition of $\hat{\chi}$. It is useful
and rather manageable even on modest computing hardware to pick around $n=1000$
sampling points and employ a higher order nonlinear optimization algorithm
such as conjugate gradient or
BFGS~\cite{fletcher2000practical} minimisation,
using gradients that most conveniently can be obtained via automatic differentiation.

This basic approach is able to produce a high-quality solution with
1000 support points on a somewhat modern laptop in less than 30 minutes.

\bibliographystyle{utphys}
\bibliography{Dil-refs-deduped.bib}

\end{document}